\documentclass[aps,pra,reprint,floatfix,longbibliography]{revtex4-1} \newcommand{\ext}{png}

\pdfoutput=1
\usepackage{amssymb}
\usepackage{mathrsfs}
\usepackage{amsmath,amsfonts,bm,graphicx,epstopdf}
\usepackage{mathtools}
\usepackage{color}
\usepackage{placeins}
\usepackage{etoolbox}
\usepackage[colorlinks=true,linkcolor=blue,urlcolor=blue,citecolor=blue]{hyperref}

\begin{document}

\newcommand{\refb}[1]{(\ref{#1})}
\newcommand{\Ref}[1]{Ref. \cite{#1}}
\newcommand{\fig}[1]{Fig.~\ref{#1}}
\newcommand{\figs}[2]{Figs.~\ref{#1} and  \ref{#2}}
\newcommand{\Fig}[1]{Figure~\ref{#1}}
\newcommand{\eq}[1]{Eq.~(\ref{#1})}
\newcommand{\eqr}[2]{Eqs.~(\ref{#1})-(\ref{#2})}
\newcommand{\Eqr}[2]{Equations~(\ref{#1})-(\ref{#2})}
\newcommand{\eqs}[2]{Eqs.~(\ref{#1}) and (\ref{#2})}
\newcommand{\Eqs}[2]{Equations~(\ref{#1})-(\ref{#2})}
\newcommand{\Eq}[1]{Equation~(\ref{#1})}
\newcommand{\App}[1]{Appendix~\ref{#1}}
\renewcommand{\sec}[1]{Sec.~\ref{#1}}
\newcommand{\Sec}[1]{Section~\ref{#1}}

\definecolor{darkred}{rgb}{0.6,0,0}
\newcommand{\red}[1]{{\textcolor{red}{#1}}}
\newcommand{\blue}[1]{{\textcolor{blue}{#1}}}
\newcommand{\cmt}[1]{{\textcolor{red}{[#1]}}}
\newcommand{\qn}[1]{{\textcolor{red}{ (?)  #1 }}}
\newcommand{\del}[1]{{\textcolor{blue}{ [deleted:  #1 ]}}}
\newcommand{\chk}[1]{{\textcolor{green}{#1}}}
\newcommand{\chg}[1]{{\textcolor{darkred}{#1}}}
\newcommand{\rvs}[1]{{#1}}

\newcommand{\e}[1]{\times 10^{#1}} 
\newcommand{\set}[1]{\{#1\}}
\newcommand{\av}[1]{\left\langle #1 \right\rangle}
\newcommand{\abs}[1]{\lvert #1 \rvert}
\newcommand{\fraci}[2]{#1/#2}

\newlength{\figwidth}
\setlength{\figwidth}{0.49\textwidth}

\newcommand{\important}[1]{{\textcolor{red}{#1}}}
\newcounter{DiscussCnt}
\newcommand{\discuss}[1]{\stepcounter{DiscussCnt}{\textcolor{blue}{[\underline{Discussion \arabic{DiscussCnt}}: #1]}}}
\newcommand{\optional}[1]{{\textcolor{darkgreen}{[\underline{Optional}: #1]}}}

\renewcommand{\r}{{{\bf r}}} 
\newcommand{\rmean}{{{\bf r}_i^{k+1/2}}} 
\newcommand{\rc}{\r}
\newcommand{\Dr}{\Delta\r}
\newcommand{\dr}{\Delta r}
\newcommand{\deltatwo}{\delta^2}

\renewcommand{\u}{{\bf \tilde r}}
\newcommand{\Den}{\rho}
\newcommand{\den}{\tilde\rho}
\newcommand{\denz}{\tilde\rho}
\newcommand{\denh}{\rho^H}
\newcommand{\rik}{\r_i^k}
\newcommand{\Drik}{\Delta \r_i^k}
\newcommand{\drik}{\Delta r_i^k}
\newcommand{\dtau}{\tau}
\newcommand{\Dtau}{\tau_{min}}
\newcommand{\taumin}{\tau_{min}}
\newcommand{\taumax}{\tau_{max}}
\newcommand{\R}{R(\dtau)}
\newcommand{\Rmax}{R(\taumax)}
\newcommand{\Rmin}{R(\taumin)}
\newcommand{\avz}[1]{\av{#1}_{i \in \Omega_z}}

\title{Deeper penetration of surface effects on particle mobility than on hopping rate in glassy polymer films}
\author{Chi-Hang Lam}
\email[Email: ]{C.H.Lam@polyu.edu.hk}
\affiliation{Department of Applied Physics, Hong Kong Polytechnic University, Hong Kong, China}
\date \today
\begin{abstract}
Free surfaces in glassy polymer films are known to induce surface mobile layers with enhanced dynamics. Using molecular dynamics simulations of a bead-spring model, we study a wide variety of layer-resolved structural and dynamical properties of polymer films equilibrated at a low temperature. Surface enhancement on thermally induced particle hopping rate is found to terminate abruptly only about 5 particle diameters from the free surface. In contrast, enhancement on the net motions of particles measured at longer time scales penetrates at least 2 particle diameters deeper. The diverse penetration depths show the existence of a peculiar sublayer, referred to as the inner-surface layer, in which surface enhanced mobility is not caused by more frequent particle hops but instead by a reduced dynamic heterogeneity associated with diminished hopping anti-correlations. Confinement effects of the free surface thus provide a unique mechanism for varying the dynamic heterogeneity and hopping correlations while keeping the hopping rate constant. Our results highlight the importance of correlations among elementary motions to glassy slowdown and suggest that dynamic facilitation is mediated via perturbations to the correlations rather than the rate of elementary motions.
\end{abstract}

%\pacs{83.80.Sg, 64.70.pj, 61.20.Ja, 47.15.gm}
\maketitle
%\tableofcontents
\section{Introduction}

The nature of glassy dynamics is a long-standing problem attracting intensive investigations \cite{donthbook,binderbook,biroli2013review,stillinger2013review,ediger2012review}.
Confinement effects on glassy thin films are widely studied in attempt to provide additional insights \cite{ediger2012review,ediger2013review,tsui2014review,napolitano2017}. 
It has already been suggested early on, based on polymer thin film experiments, that  surface layers with enhanced dynamics dominate thin film confinement effects \cite{keddie1994,kawana2001}. This is supported by molecular dynamics (MD) simulations \cite{baschnagel2006,baschnagel2007} and has been further established by more recent film flow experiments \cite{tsui2010,ediger2011,forrest2014}. In particular, our experiments on short-chain polymer thin films in Ref. \cite{tsui2010} have shown that a thin surface mobile layer exists on top of a glassy bulk layer. The dynamics of the surface layer follow an Arrhenius temperature dependence. This non-glassy nature of the surface layer is consistent with indications from earlier experiments \cite{tanaka1998,forrest2008}. 

A main motivation of the study by Keddie et al. on polymer films was to cast light onto the fundamental origins of glass \cite{keddie1994}. This has proved challenging due to complications including substrate influences and possible long-range elastic couplings by long polymer chains. Short-chain polymer films on supposedly non-slipping and non-permanently pinning substrates such as those used in \Ref{tsui2010} may thus provide the simplest scenario. 
Nevertheless, the origin and the detailed properties of the surface mobile layer are still controversial \cite{herminghaus2002,long2001,lipson2010,starr2014,forrest2015string}. 
For simplicity, one often assume  
simple layer models in which both the surface mobile layer and the inner glassy layer have uniform properties separated by an abrupt boundary \cite{keddie1994,kawana2001,tsui2010,paeng2011,tsui2013pmma}. However, films are expected to have graded depth-dependent properties \cite{roth2010,ogieglo2018,mckenzie2018} as found in MD simulations \cite{jain2004,baschnagel2006,baschnagel2007,starr2012,lam2013crossover}.

It is long known from MD simulations that perturbations due to a free surface of a polymer film penetrate much deeper for dynamical than for structural properties \cite{baschnagel2006,baschnagel2007,lam2013crossover}.
However, different penetration depths for different dynamical measurements of relevance would be unexpected and have not been identified in our knowledge.

In this work, we perform large-scale MD simulations of polymer films at equilibrium at \rvs{an exceptionally low temperature and a zero pressure} using GPU-based brute-force computing with individual runs executed for 5 months. Comprehensive measurements on depth-dependent structural and dynamical properties are performed. Surface enhanced dynamics is exemplified in particular by a higher particle hopping rate close to the surface. Unexpectedly, the surface enhancement on the hopping rate terminates abruptly when going deeper into the film. This is similar for other hop related dynamical measurements at short time scales.
In contrast, surface effects on other dynamical measurements at longer time scales show much deeper penetrations. They include particle mobility measured at longer time scales and hopping event correlations quantifying dynamic heterogeneity. There thus exists a region in which the particle hopping rate is bulk-like but the particle mobility and dynamic heterogeneity are surface affected. This is a unique example in which one can perturb certain dynamic quantities relevant to structural relaxations while maintaining other dynamic quantities unchanged.

The rest of the paper is organized as follows. \Sec{method} introduces the model and the simulation methods. Sections \ref{static} and \ref{dynamics} present results on structural and dynamical properties, which motivate the definition of three sublayers of the surface mobile layer. We then discuss in \sec{sublayer} detailed properties and possible origins of the sublayers. Implications of our findings on theoretical understanding of glass is discussed in \sec{facil}. Finally, \sec{discussions} concludes the paper with a summary and some further discussions.

\section{Model and simulation methods}
\label{method}

Our simulations are based on the Kremer-Grest model of bead-spring polymer widely used in the literature \cite{kremer1990,varnik2002pre,varnik2002,kremer2003,scheidler2004,baljon2005,doi2006,baschnagel2006,baschnagel2007,starr2012,lam2013crossover,lam2017}. We adopt the variant used in \Ref{lam2017} in which polymer chains possess heavier chain-tails so that all monomers have similar mobilities.
Specifically, we simulate polymer melts consisting of chains which are 10 monomers long. \rvs{This is well below the entanglement chain length \cite{hou2010} and it thus models unentangled short-chain polymer following Rouse dynamics \cite{edwardsbook}.}
We refer to the monomers as particles.
Pairs of particles interact via the Lennard Jones (LJ) potential 
$4\epsilon \left[\left({\sigma}/{r}\right)^{12} - \left({\sigma}/{r}\right)^{6} \right]$
with an interaction cutoff distance $R_c = 2\cdot2^{1/6}\sigma \simeq 2.24\sigma$ beyond which it becomes a constant. 
Besides short-range repulsion, the potential implements longer-range attraction which is essential for simulating polymer films with free surfaces. 
Bonded particles are further bounded by a finitely extensible nonlinear elastic (FENE) potential 
$ - \frac{k}{2} R_0^2 \ln \left[ 1 -\left({r}/{R_0}\right)^2 \right]$
where $k=30\epsilon/\sigma^2$ and $R_0=1.5\sigma$. 
We adopt dimensionless LJ units which amounts to taking $\sigma=\epsilon=1$.
Internal particles in each chain have a mass $M_I=1$. The heavier chain-tail property amounts to assigning a larger mass $M_T=4$ to the particles at both ends of each chain. This leads to an approximately uniform mobility for all particles as indicated by particle mean squared displacements (MSD). 
Throughout the paper, we express lengths in unit of $\sigma$ for clarity even though  $\sigma \equiv 1$.

Simulations are performed in a box of dimensions $L \times L \times \infty$ with $L=24\sigma$ following periodic boundary conditions in the $x$ and $y$ directions. We consider free-standing polymer films each having 1200 chains leading to totally 12000 particles. This provides free surfaces on both sides. Centering the film at $z=0$, results presented for $z\ge 0$ in general are averages over the $\pm z$ positions.

\rvs{Our main results are measured from films fully equilibrated at temperature $T=0.36$, which is about the lowest temperature for practical equilibrium simulation of the model with existing computing technologies. The pressure is zero since empty regions exist above and below the film.} All simulations are performed using the HOOMD software package \cite{hoomd} under NVT conditions with a time step of 0.005.  The chain configurations in the films are initialized randomly and  thermalized at $T=0.8$ following standard techniques \cite{varnik2002pre,kremer2003}.
They are subsequently annealed at $T=0.5$ and then repeatedly cooled and annealed by temperature steps of $\Delta T=0.02$. Individual cooling and annealing processes both  involve $10^7$ timesteps. At the final $T=0.36$, each film is further annealed for $10^{10}$ timesteps before data taking. This procedure has been found to generate well-equilibrated polymer melts in bulk simulations with particle MSD exceeding $5.0\sigma^2$ \cite{lam2017}. The equilibration takes about 5 months using a nVidia GTX580 GPU. 

We prepare 5 independent film samples using the procedures above.
For each sample, we perform 3 data collection runs at $T=0.36$ each of $8\e7$ timesteps. To enhance the statistics, these data collection runs are separated by further annealing of $10^9$ timesteps during which data are not collected. All our quantitative measurements are averaged over both  surfaces in these 15 data collection runs. Errors are estimated from fluctuations among the 5 independent film samples. The computations involved in this study are in our knowledge the most intensive ones for bead-spring polymer films reported in the literature.

All our analysis are based on coarse-grained particle trajectories $\rc_i(t)$ 
recorded during data collection runs. Here, each value of $\rc_i(t)$ is a coarse-grained position of particle $i$ defined by
\begin{equation}
  \rc_i(t) = \langle ~ \r^0_i(t') ~ \rangle _ {t' \in [t, t+\Delta t_c] }
\end{equation}
where $\r^0_i(t)$ denotes its instantaneous position. The averaging duration $\Delta t_c= 5$ is chosen to be well in between the typical particle vibrational oscillation period and the  waiting time between two consecutive hops of a particle. Therefore, $\rc_i(t)$ nearly always points to a meta-stable particle position, rather than somewhere interpolating between two meta-stable positions related by a hop. In each data collection run, we record coarse-grained trajectories in the form of $2\e4$ snapshots of coarse-grained positions $\rc_i(t)$ taken after every 4000 timesteps corresponding to a duration of $\taumin=20$. 

\section{Static structural properties}
\label{static}

\begin{figure}[tb] %%%%%%%%%%%%
\includegraphics[width=\figwidth]{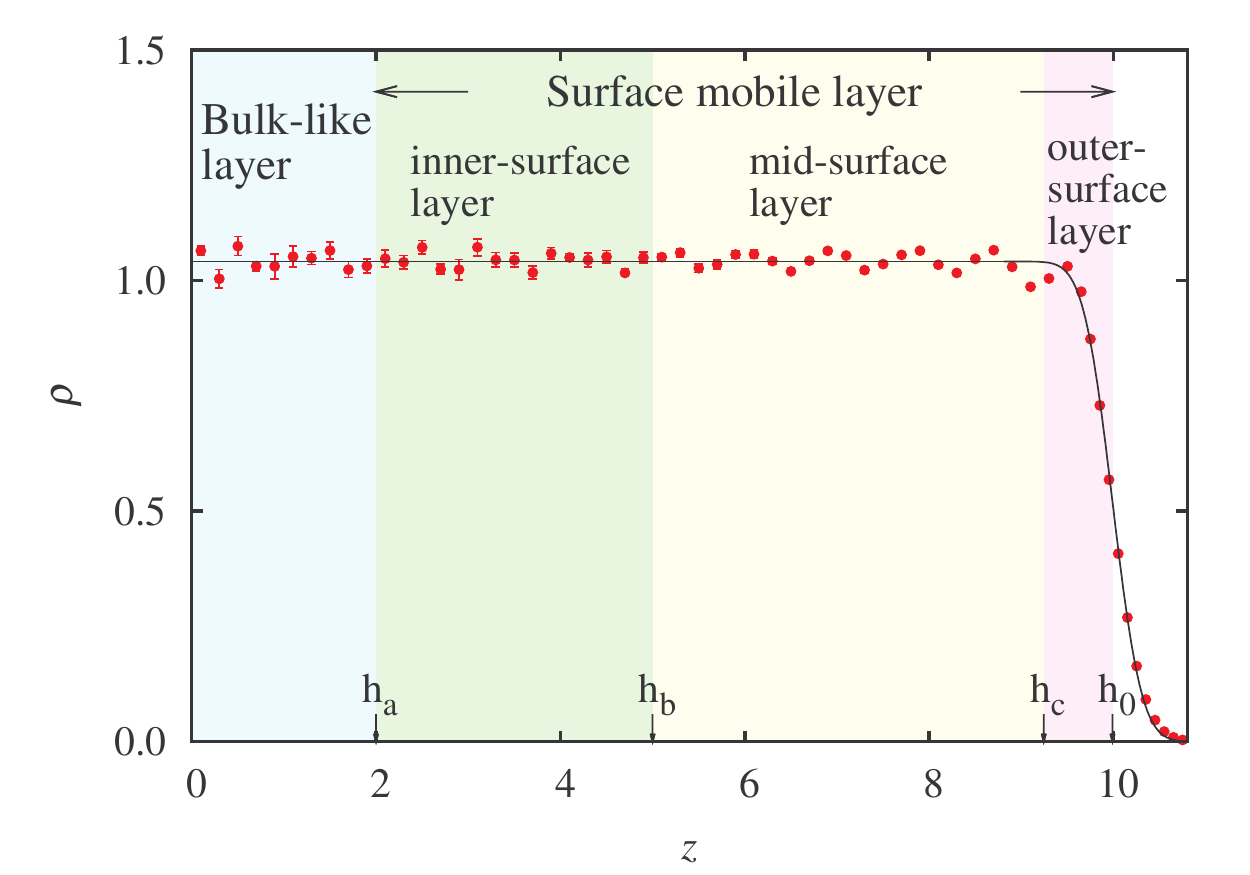}
\caption{(a) Particle density $\Den$ against coordinate $z$ for free-standing films   centered at $z=0$. The solid curve represents a fit to \eq{rhoz}.  The position $z=h_0 \equiv 9.99 \sigma$, where ${\sigma\equiv 1}$, marks the average position of the film surface, with $h_0$ being the half film thickness. The surface mobile layer consists of the outer-, mid- and inner-surface layers, which are bounded below respectively by 
$h_a \equiv 2.0\sigma$, $h_b \equiv 5.0\sigma$ and $h_c\equiv 9.25\sigma$.
}
\label{Fdensity}
\end{figure}

\Fig{Fdensity} plots the particle density $\Den$ of the film as a function of the non-planar coordinate $z$. 
It shows that $\Den$ converges to its bulk value at a very shallow depth from the free surface as observed in previous works \cite{baschnagel2007,lam2013crossover}. It is well fitted by \cite{lam2013crossover}
\begin{equation}
  \label{rhoz}
  \Den = \frac{\rho_0}{2} \text{Erfc}\left( \frac{z-h_0}{\sqrt{2} \sigma_h } \right) 
\end{equation}
appropriate for surfaces limited by surface tension \cite{nelson2004}, where Erfc is the complementary error function. 
From the fit, we find a bulk particle density $\rho_0=1.04$, a half film thickness $h_0 = 9.99 \sigma$ and a surface width $\sigma_h = 0.248 \sigma$.
\Fig{Fdensity} also shows spatial density oscillations of a small amplitude close to the surface. This corresponds to slight layering effects and only occurs to our samples at very low $T$ after long annealing. These small modulations however appear to have negligible impacts on other properties to be discussed.

To define further layer-resolved quantities,
let $\Omega_z$ be a layer of particles in between $z\pm \Delta z/2$. In all following quantitative measurements, we consider a layer thickness $\Delta z=0.5$.
For any 3D position $\r$, the projection onto the $xy$-plane is denoted by $\u$, and it is similar for other position vectors. 
The 2D local particle density $\denz(\u)$ of layer $\Omega_z$ at 2D position $\u$ is 
then
\begin{equation}
  \label{den}
  \denz(\u) = \sum _ {i\in\Omega_z} \deltatwo (\u - \u_i )
\end{equation}
where $\deltatwo$ denotes the 2D Dirac delta function. 
The 3D particle density $\Den$ shown in \fig{Fdensity} relates to $\denz$ by ${\Den = \av{\denz(\u_0)}/\Delta z}$,
where the average is taken over all 2D positions $\u_0$.

\begin{figure}[tb] %%%%%%%%%%%%
\includegraphics[width=\figwidth]{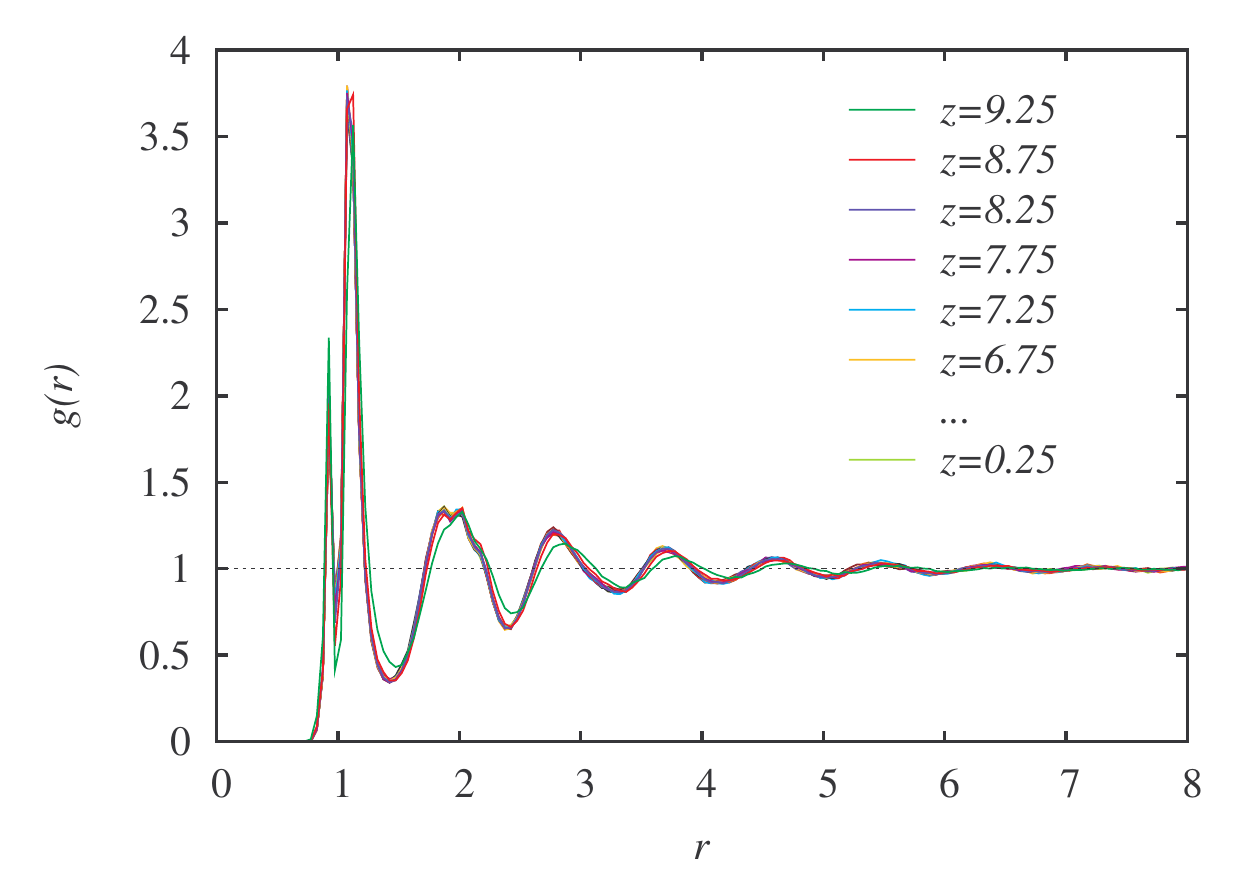}
\caption{(a) 
2D pair distribution function of particles $g(r)$ against 2D distance $r$ defined on the $xy$ plane. 
All curves collapse apart from some slight deviations for $z=9.25\sigma$.
}
\label{Fdencorr}
\end{figure}

The 2D pair distribution function of particles in layer $\Omega_z$ can now be defined as
\begin{equation}
  \label{gr}
  g(r) = \frac{ \av{ \denz( \u_0) \denz( \u_0 + \u)} }{\av{\den(\u_0)}^2}
\end{equation}
where $r= \abs \u$ and averages are performed over 2D positions $\u_0$. 
In practice, it is evaluated using the equivalent form
\cite{hansen1990book} % page 30
\begin{equation}
  \label{gr2}
  g(r) = \frac{1}{\pi r L^2 \av{\den(\u_0)}^2} \av{\sum_{\substack{~i,j\in \Omega_z\\(i>j)}} \deltatwo (\u- \abs{\u_i-\u_j} )}.
\end{equation}
\Fig{Fdencorr} plots $g(r)$ against $r$ for various layer positions $z$. 
The main peak at $r\simeq \sigma$ is split into two subpeaks corresponding to bonded and non-bonded nearest neighbors, while the weaker peaks are due to further neighbors.
Moreover, $g(\u)$ for $z\alt 9.25\sigma$ are practically independent of $z$, representing the bulk values. Deviations dramatically increase only for $z\agt 9.75\sigma$ (data not shown).

    Therefore, both $\Den$ and $g(r)$ exhibit bulk-like values except at very close to the free surface. Surface effects on structural measurements studied are significant only for $z \ge h_c \equiv h_0-3\sigma_h = 9.25 \sigma$ and
we refer to the region as the outer-surface layer (see \fig{Fdensity}). 

\section{dynamical properties}
\label{dynamics}

\subsection{Displacement statistics}
\label{displacement}

Let $\dr_i$ be the displacement of particle $i$ over a duration $\dtau$ at time $t_0$ defined by
\begin{equation}
  \label{dri}
  \dr_i = { \mid\r_i(t_0+\dtau)- \r_i(t_0) \mid }, 
\end{equation}
which is a net displacement in general shorter than the distance traveled along the actual path.
The particle MSD at layer $\Omega_z$ % as a function of duration $\dtau$ 
is then given by
\begin{equation}
  \text{MSD} = \avz{ \dr_i ^2 }
  \label{MSD}
\end{equation}
where the average is limited to particles in the layer $\Omega_z$. Particle $i$ is deemed inside $\Omega_z$ if its position $\r_i(t_0)$ at the initial time $t_0$ of the displacement is within the layer. Since we focus on the small displacement regime, adopting other more stringent criteria \cite{baschnagel2007} does not alter our results qualitatively and this is further discussed in \App{layercriteria}. 

\begin{figure}[tb] %%%%%%%%%%%%
\includegraphics[width=\figwidth]{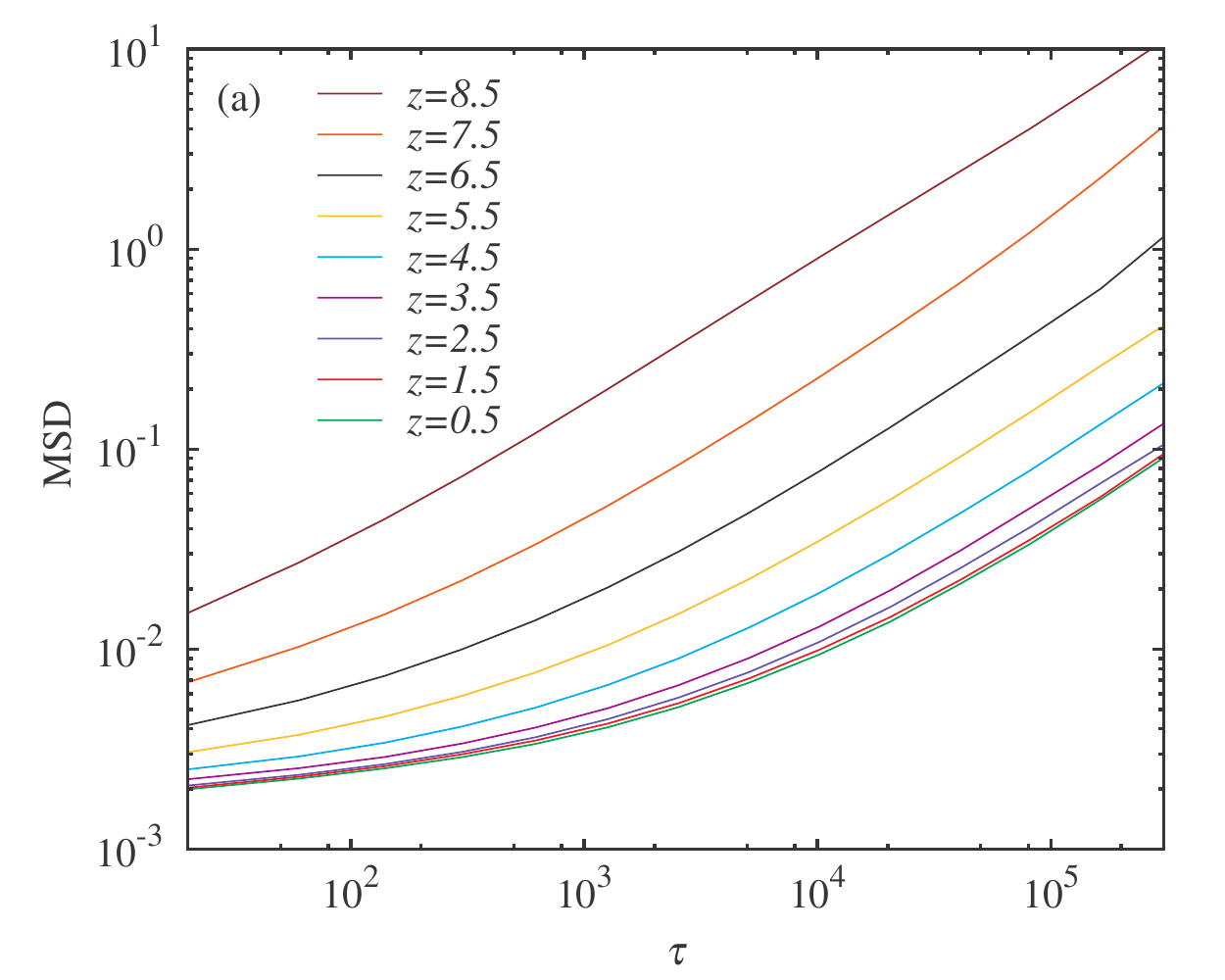}
\includegraphics[width=\figwidth]{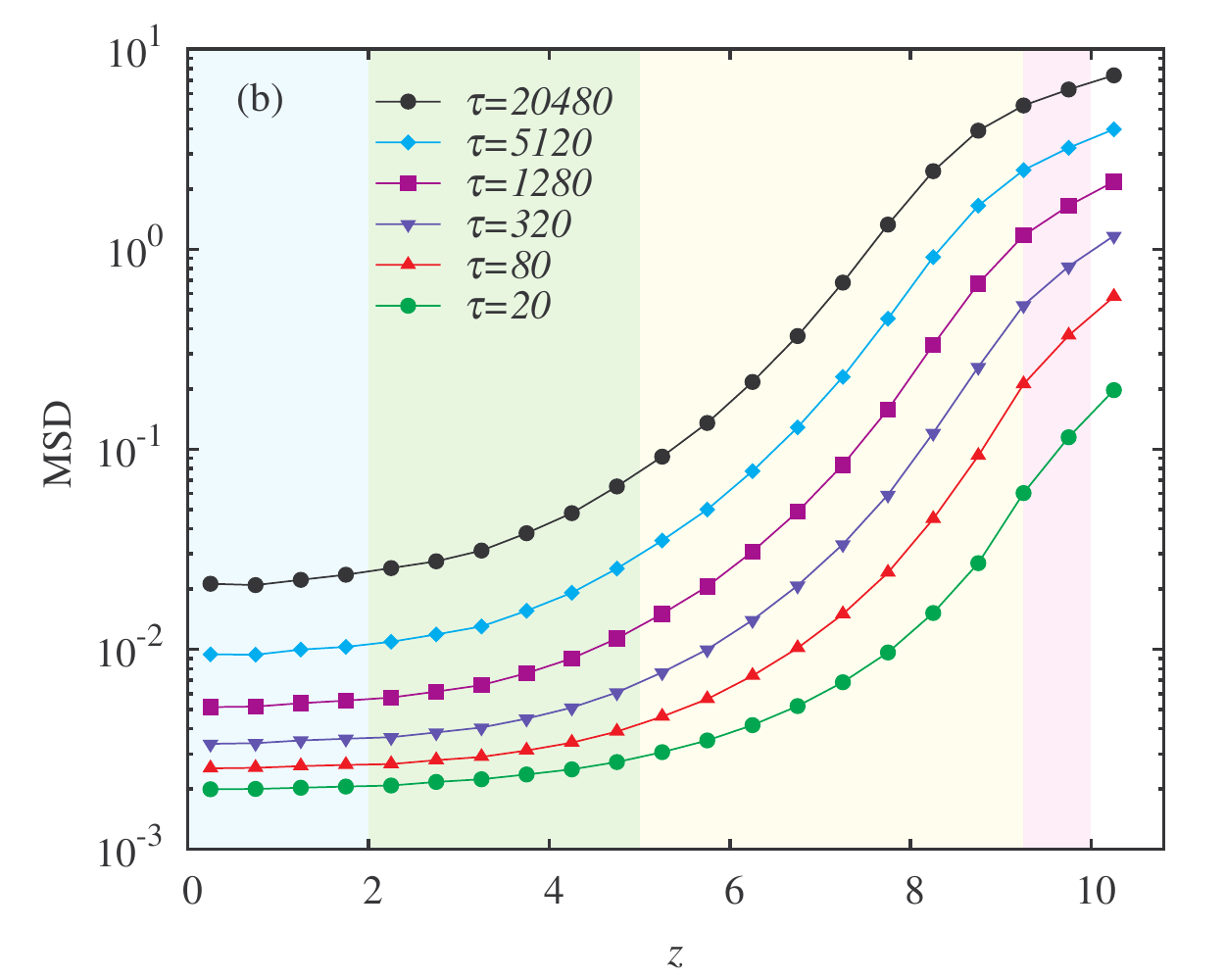}
\caption{ (a) Particle mean square displacement (MSD) against duration $\tau$ for $z = 0.5\sigma, 1.5\sigma, \dots$ (from bottom to top) where $\sigma\equiv 1$. The MSD increases dramatically with $z$ for $z\agt 3.5\sigma$ indicating surface enhanced mobility.
(b) MSD against coordinate $z$ plotted using the same data as in (a). 
%From (b), surface regions with enhanced  MSD enlarges as time $\tau$ %increases.
}
\label{Fmsd}
\end{figure}

\Fig{Fmsd}(a) plots the measured MSD against the duration $\dtau$ for various layer positions $z$. The layer thickness used is $\Delta z=0.5$ but we have performed averaging over every two neighboring layers to thin out the data for clarity. Results are qualitatively similar to those in \Ref{baschnagel2007}. 
For $z\alt 2.5$, the MSD is approximately independent of $z$ representing the bulk values. In contrast, for $z\agt 3.5\sigma$, it increases dramatically with $z$, demonstrating surface enhanced dynamics. This also shows that surface effects penetrate far beyond the outer-surface layer and extend much deeper than those on the structural properties as observed previously \cite{baschnagel2007}. This is more clearly observed in \Fig{Fmsd}(b) which replots the MSD against $z$ for various duration $\dtau$. It is also evident that surface effects penetrate deeper as $\dtau$ increases. 

\begin{figure}[tb] %%%%%%%%%%%%
\includegraphics[width=\figwidth]{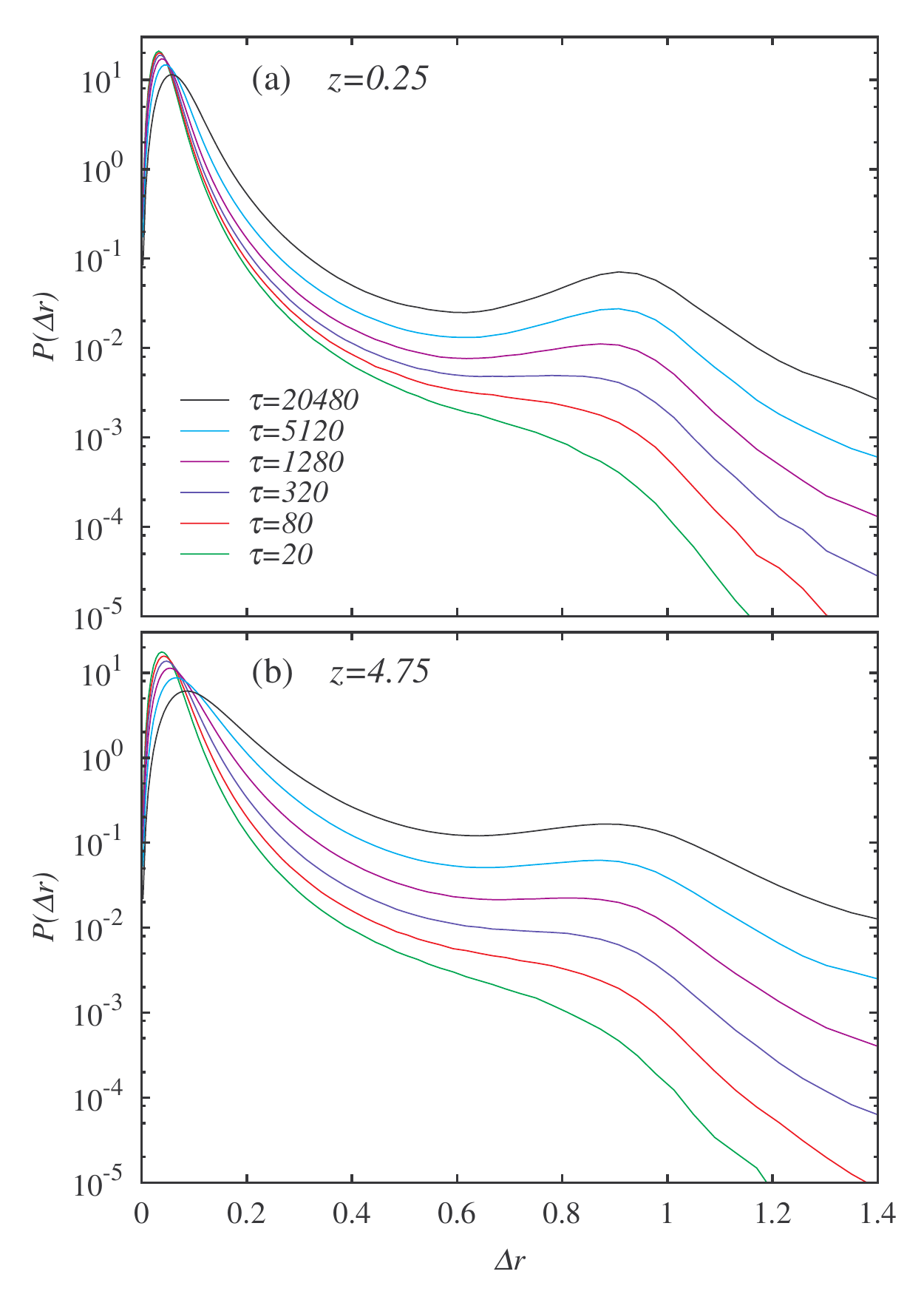}
\caption{Probability distribution $P(\dr)$ of particle displacement $\dr$ over  duration $\dtau$ at $z=0.25\sigma$ in the bulk-like layer (a) and at $z=4.75\sigma$ in the inner-surface layer (b). In both cases, the main peak at $z\simeq 0$ corresponds to particles which have not hopped and its broadening at increasing  $\dtau$ is due to particle creep motions. The secondary peak at $z\simeq 0.9\sigma$  is due to particle hops. The dip at $z\simeq 0.6\sigma$ is adopted as the threshold of hopping. 
}
\label{FPr1}
\end{figure}

\begin{figure}[tb] %%%%%%%%%%%%
\includegraphics[width=\figwidth]{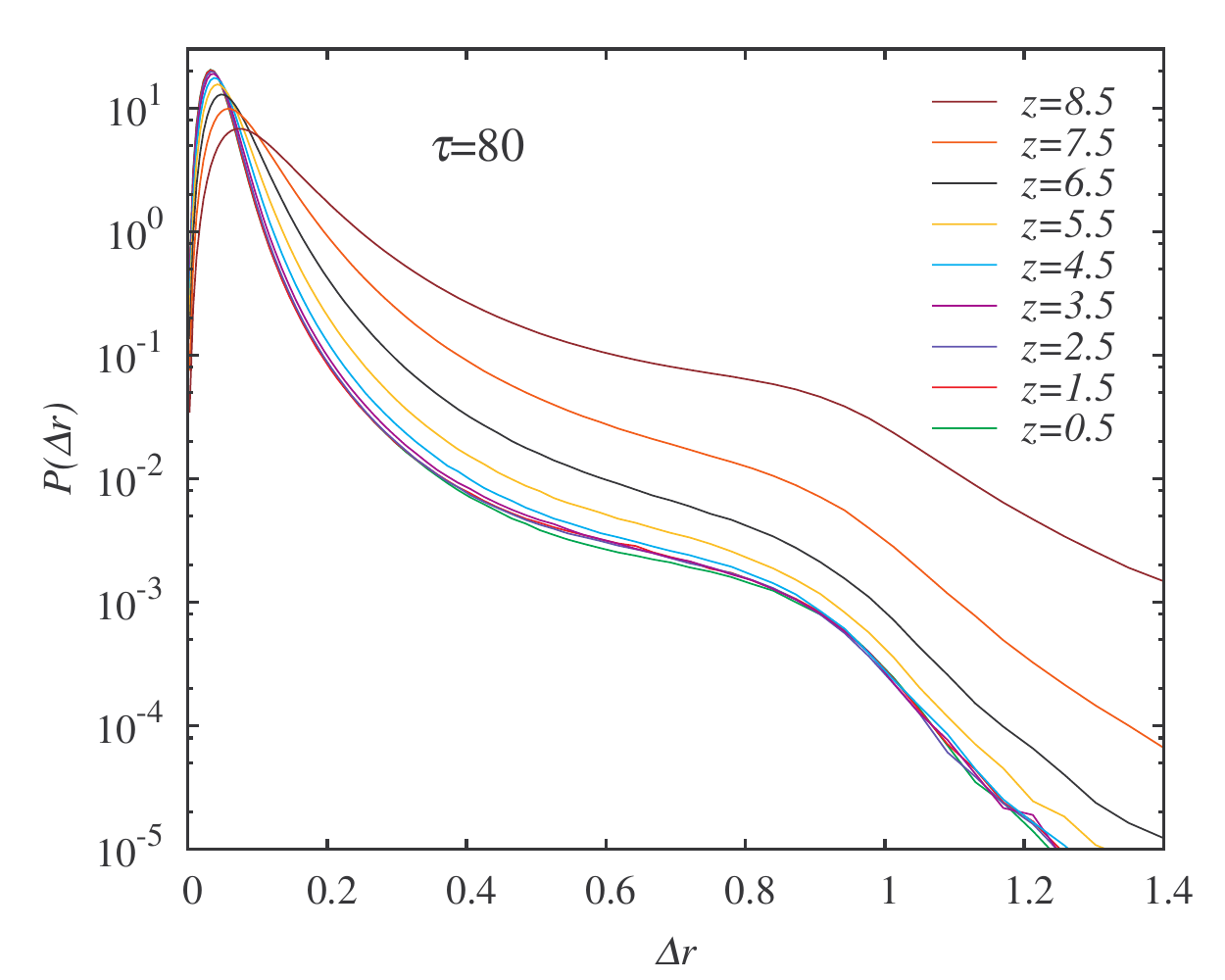}
\caption{ Probability distribution $P(\dr)$ of particle displacement $\dr$ over a short duration $\tau=80$. At $z=4.5\sigma$, it begins to deviate noticeably from its bulk value signifying the onset of surface enhanced motions. The deviation is 
proportionately more prominent at $\Dr \simeq 0.2\sigma$ than at $0.9 \sigma$,  implying a stronger enhancement on particle creep motions than on hops.
}
\label{FPr80}
\end{figure}

We next study the probability distribution $P(\dr)$ followed by the displacement $\dr_i$, which  
is closely related to the van Hove self-correlation function \cite{wahnstrom1991}.
\Fig{FPr1}(a) plots the computed $P(\dr)$ at ${z=0.25\sigma}$ deep in the bulk-like region for various duration $\dtau$. The results are similar to those from bulk simulations of the same model in \Ref{lam2017}. 
Besides a main peak, a secondary peak at ${\dr \simeq 0.9 \sigma}$ emerges as $\dtau$ increases and  corresponds to particle hops. The activated nature of hopping is  evidenced by a dip in $P(\dr)$ at ${\dr \simeq 0.6 \sigma}$. 

Besides hops, let us refer to all other non-vibrational motions mainly due to accumulation of smaller displacements as particle creep motions. At small duration $\dtau$, a displacement $\dr > 0.6 \sigma$ thus usually results from a hop while $\dr < 0.6 \sigma$ usually implies vibrations or creep motions.
From \fig{FPr1}(a), as $\dtau$ increases from 20 to 80, $P(\dr)$ at $\dr \simeq 0.9\sigma$ increases considerably and indicates significant hopping motions. The main peak however broadens only slightly implying that creep motions are negligible.

\Fig{FPr1}(b) shows another example of $P(\dr)$ at ${z=4.75\sigma}$ which admits some mild surface enhanced dynamics as indicated by the MSD in \fig{Fmsd}(a). This choice of $z$ will be more apparent later.
Results in \fig{FPr1}(b) are similar to those in \fig{FPr1}(a) except that as 
$\dtau$ increases from 20 to 80, the broadening of the main peak is much more significant. This indicates that the surface enhanced dynamics at $z=4.75\sigma$ are contributed significantly by  enhanced creep motions. 

To better compare between the layers, \fig{FPr80} plots $P(\dr)$ at $\dtau=80$ for various $z$. For $z \ge 5.5\sigma$, $P(\dr)$ is beyond the bulk value for all $\dr$. At $z = 4.5\sigma$ close to the onset of surface effects, $P(\dr)$ is slightly but distinctly beyond the bulk value for $\dr \simeq 0.3 \sigma$ indicating broadening of the main peak. However, it is indistinguishable from the bulk value at  $\dr \simeq 0.9 \sigma$ showing no growth of the secondary peak.  This hence shows that the surface  effects reach as deep as $z=4.5\sigma$ for creep motions but not for hops. An explanation of this observation will be discussed in \sec{sublayer}.

\subsection{Particle hopping rates}
\label{hoppingrates}

\begin{figure}[tb] %%%%%%%%%%%%
\includegraphics[width=\figwidth]{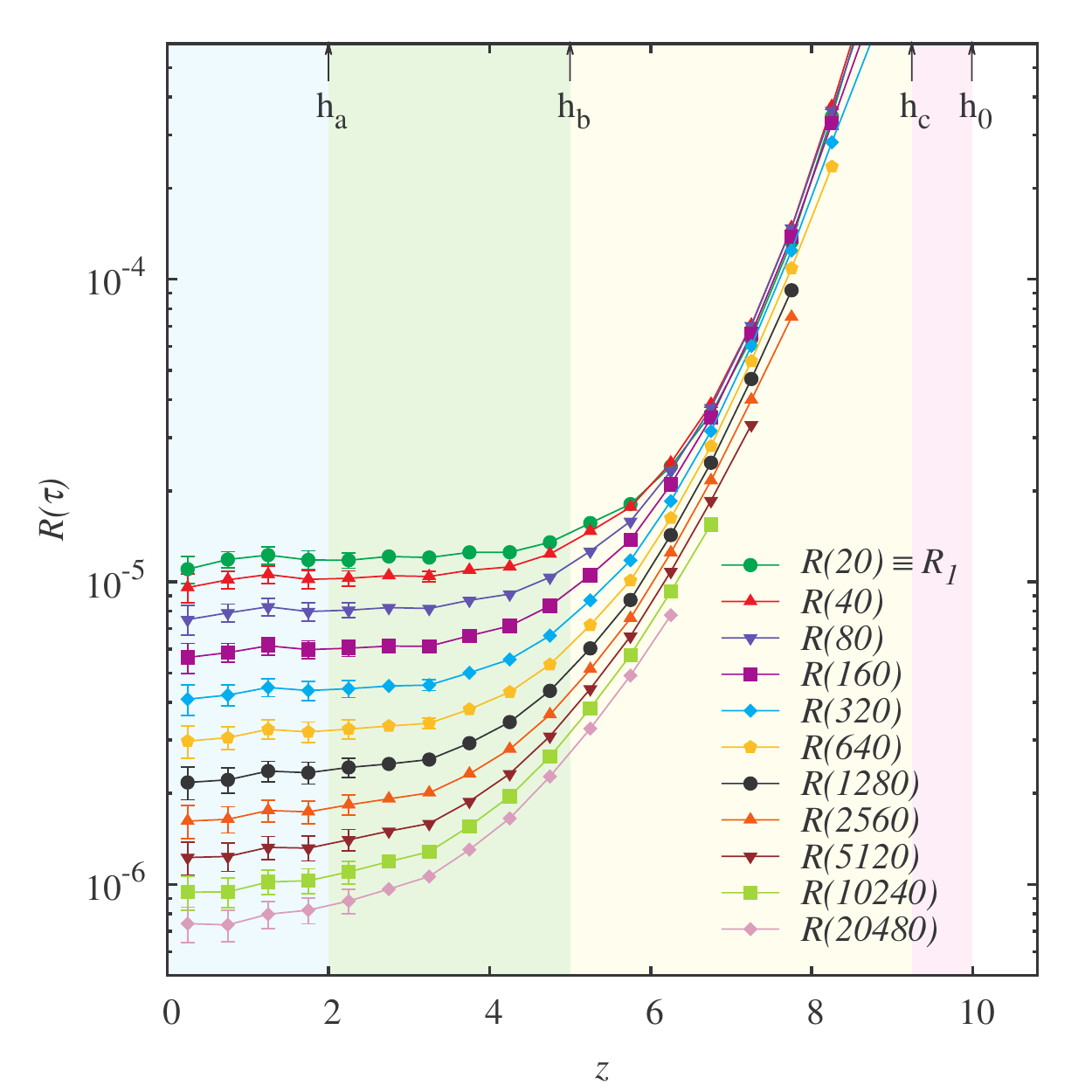}
\caption{ Particle net hopping rate $\R$ against coordinate $z$ based on net displacement over duration $\dtau$. Particle hopping rate $R_1$ is identified as $\Rmin$ where $\taumin=20$. Since $R_1$ converges to is bulk value quite abruptly at $z\simeq h_b\equiv 5\sigma$, surface effects on $R_1$ do not reach the inner surface layer. In contrast, $R(\tau$) at larger $\tau$ converges only deeper into the film, indicating that their surface effects penetrate into the inner surface layer as well.  Errorbars smaller than the symbols are omitted.
}
\label{FR}
\end{figure}

We considered particle $i$ as having hopped during a period $\dtau$ if its displacement $\dr_i$ defined in \eq{dri} satisfies
\begin{equation}
  \label{hop}
    \dr_i ~ \ge ~ 0.6 \sigma
\end{equation}
following \Ref{lam2017}, 
where the threshold $0.6 \sigma$ is the position of the first dip in $P(\dr)$ from \fig{FPr1}(a).
The net hopping rate $\R$ can be defined as
\begin{equation}
  \label{R}
  \R = \frac{1}{\dtau}\avz{ ~\theta ( \dr_i  - 0.6 \sigma)~}
\end{equation}
where $\theta$ denotes the Heaviside step function. This provides a net hopping rate because $\dr_i$ defined in \eq{dri} is a net displacement so that round trips, for example,  do not contribute.

\begin{figure}[tb] %%%%%%%%%%%%
\includegraphics[width=\figwidth]{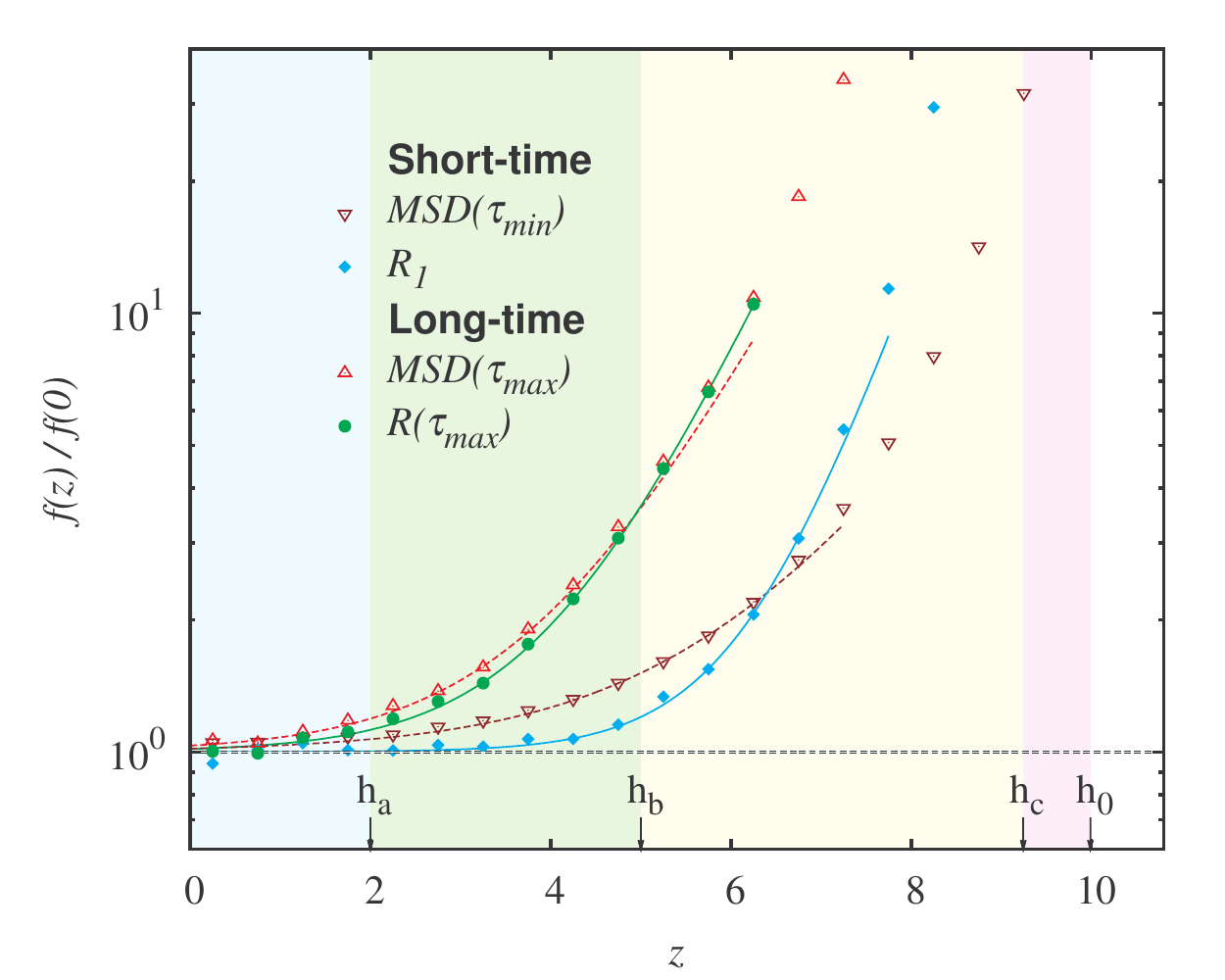}
\caption{ Layer-resolved quantity $f(z)$ normalized with respect to bulk value $f(0)$ against coordinate $z$. Quantities considered are the MSD and the particle hopping rate $R_1$ defined at short time $\taumin=20$, as well as  the MSD and the net hopping rate $\Rmax$ at long time $\taumax=20480$. 
The solid lines show fitted curves using \eq{fit}.
}
\label{FmsdR1}
\end{figure}

\Fig{FR} shows the main result of this work which is a plot of the computed $R(\dtau)$ against $z$ for various duration $\dtau$.
We have included only data satisfying $ \dtau \R \le 0.2$, a typical constraint for simple rate measurement. 
We have also checked that the displacement distribution $P(\dr)$ corresponding to each data point exhibits a clear secondary bump or peak at $0.9\sigma$ so that particle hops indeed dominate. 
From \fig{FR}, $\R$ decreases dramatically with $\dtau$ in general. 
Detailed examinations of individual particle trajectories show that this is due to the abundance of back-and-forth hopping motions at low $T$ widely studied in the literature
\cite{miyagawa1988,vollmayr2004,vogel2008,kawasaki2013,ahn2013,helfferich2014,yu2017,lu2016,lam2017}. Since $\R$ is a net hopping rate not registering the back-and-forth parts of the hops, at large $\dtau$, it underestimates the true hopping rate and is instead a better indicator of particle mobility describing long-time motions in the diffusive regime. 
Now, we approximate the true hopping rate $R_1$ using the net rate $\R$ at the smallest studied $\dtau$, i.e. 
\begin{equation}
  \label{R1}
  R_1 = R(\taumin) 
\end{equation}
where $\taumin \equiv 20$. Since $\taumin$ must also be much longer than the duration of the course of a hop, called the instanton time \cite{chandler2011}, the current value should already be about the smallest practical one. 

A main observation in this work is that $R_1$ in \fig{FR} converges rather abruptly to its bulk value exhibiting a surprisingly wide plateau which begins to curve up only not far away from the free surface. 
This is in contrast to the case of the MSD for the same duration $\taumin$ shown in \fig{Fmsd}(b). 
\Fig{FR} also shows that surface effects on $\R$ 
penetrates deeper as $\dtau$ increases, analogous to that revealed by the MSD. 

\Fig{FmsdR1} shows on the same plot both the MSD and $\R$ at $\dtau$ being $\taumin \equiv 20$ and $\taumax \equiv 20400$ 
normalized by the respective bulk values. For $\taumin$, it is clear that $R_1 \equiv \Rmin$ converges to its bulk value much more abruptly than the MSD. For $\taumax$ related to particle mobility, the normalized MSD and $\Rmax$ are close to each other.  
Fits of these quantities to an exponential functional form will be explained in \sec{sublayer}.

From \fig{FmsdR1}, since surface effects on $R_1$ are small at  
$z <h_b$ where $h_b = 5.0\sigma$, we refer to the region $h_b \le z \le h_c$ as the mid-surface layer (see \fig{Fdensity}). This sublayer is characterized by a surface enhanced $R_1$ despite bulk-like structural properties. 
Similarly, surface effects on $\Rmax$ are negligible for $z < h_a$ where $h_a = 2.0\sigma$ and we define the inner-surface layer as the region $ h_a \le z \le h_b$ (see also \fig{Fdensity}).
It is characterized by a surface enhanced $\Rmax$ despite a bulk-like $R_1$.
Note that a larger $\taumax$ may increase $h_a$, but should not affect our conclusions qualitatively. 

\subsection{Dynamic heterogeneity}

\begin{figure}[tb] %%%%%%%%%%%%
\includegraphics[width=\figwidth]{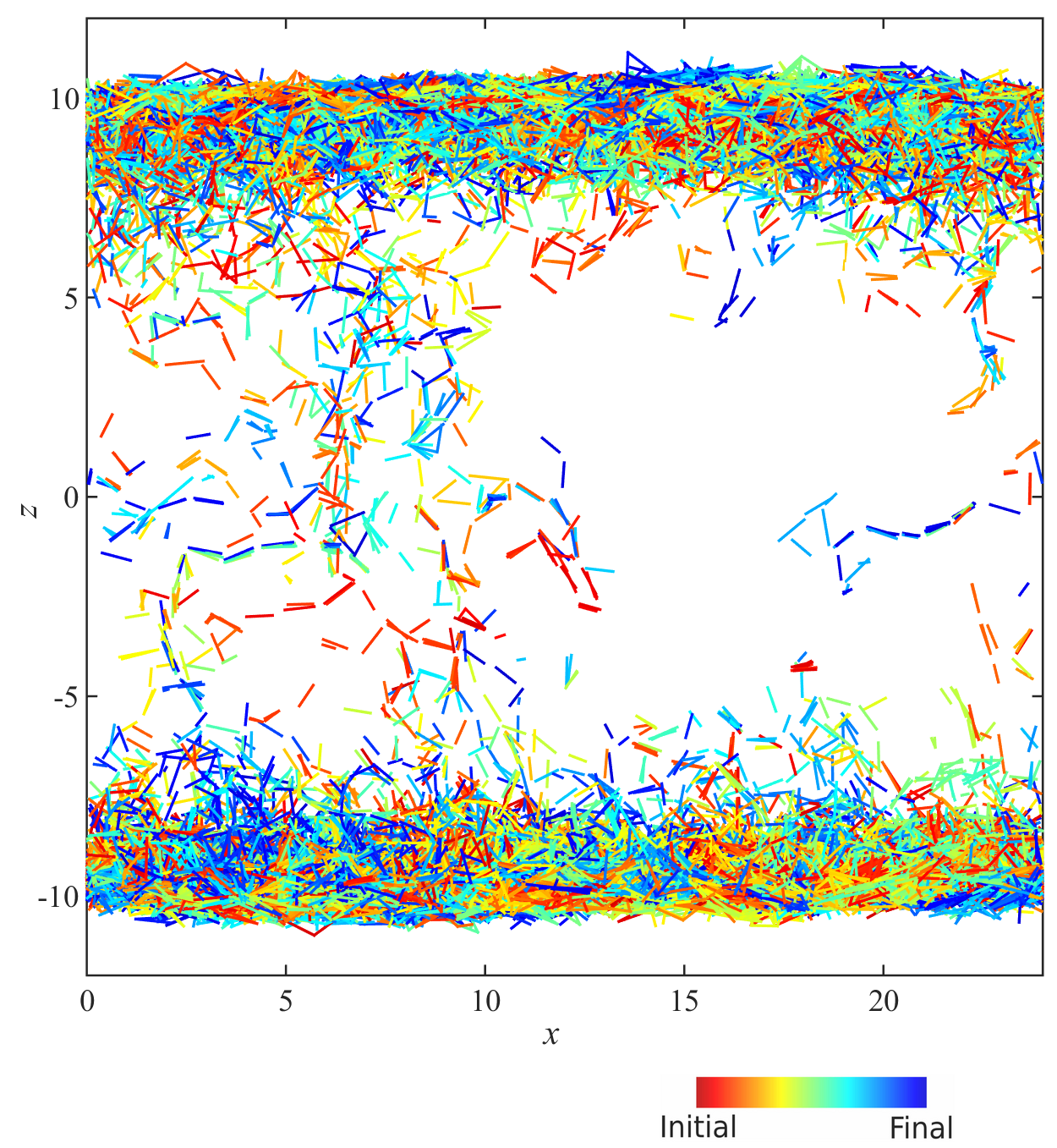}
\caption{
Hopping events as illustrated by coarse-grained particle trajectories during hopping in a typical film. 
A particle generates a hopping event if its displacement during any time interval $\Dtau=20$ is beyond $0.6\sigma$. The diagram shows all hopping events occurring over a period $\tau = 10,000$.
Color represents time at which the hop occurs relative to the duration $\tau$, as indicated in the legend. The high density of hopping events on both the upper and lower surfaces of the free standing film illustrates enhanced surface mobility. Closer to the center of the film, events are fewer and string-like motions and their repetitions can be observed. Dynamic heterogeneity is also revealed from the spatial distribution of these dynamical events. 
}
\label{Fblockfull}
\end{figure}

\begin{figure}[tb] %%%%%%%%%%%%
\includegraphics[width=\figwidth]{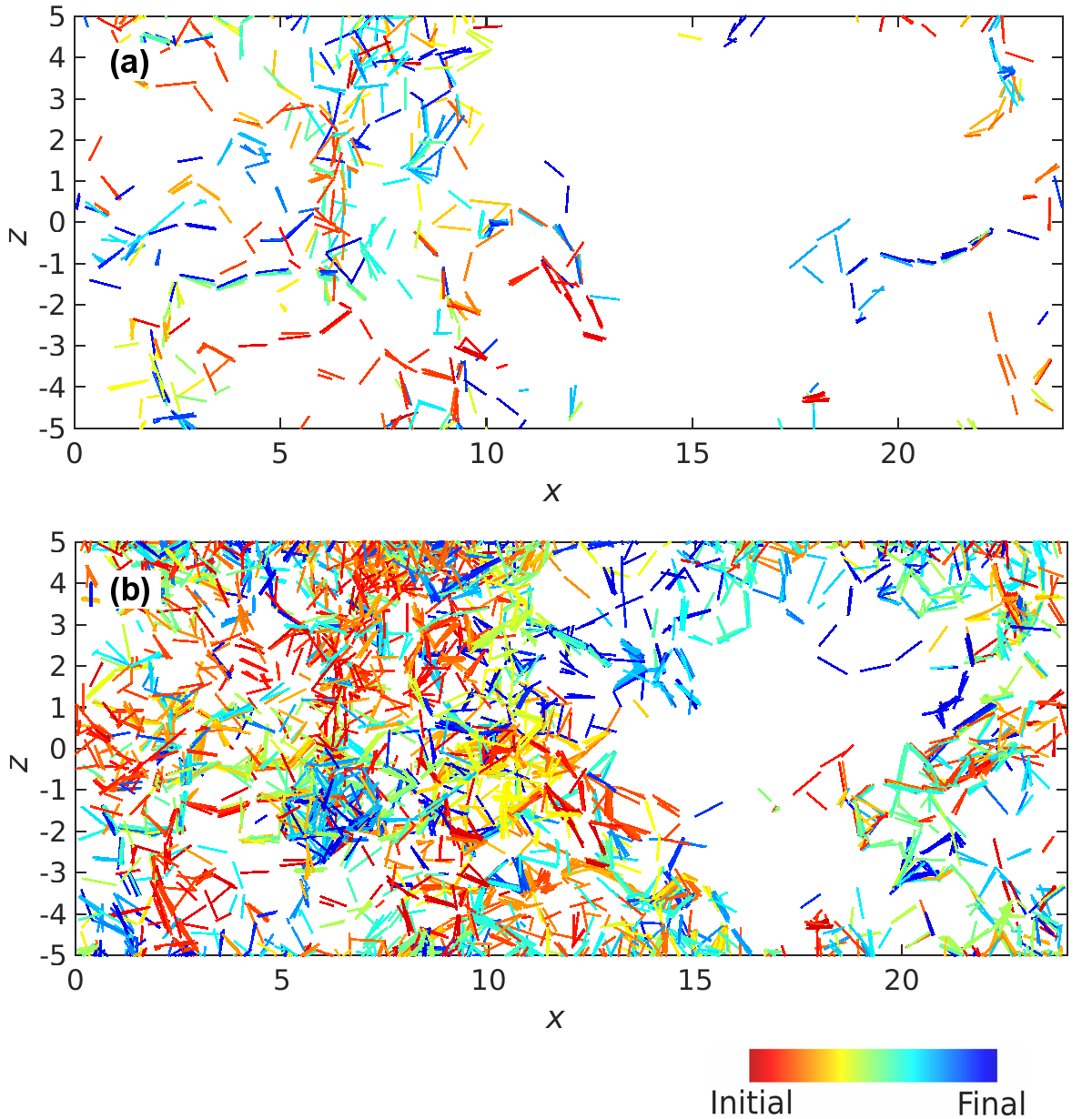}
\caption{ (a) Hopping events from   \fig{Fblockfull} but limited only to the bulk-like and inner-surface layers at $|z| \le 5.0\sigma$ both following the bulk-like hopping rate $R_1$. Events occurring over a period $\tau = 10,000$ are shown. (b) The period is extended to $\tau=100,000$ for the same polymer sample. A stronger dynamic heterogeneity is  observed closer to the center than at the margins of the region, although the density of the hopping events is statistically   uniform.  This visually illustrates the co-existence of a $z$-independent average particle hopping rate and a $z$-dependent  dynamic heterogeneity. A similar but weaker trend concerning the dynamic heterogeneity can also be observed in (a).
}
\label{Fblock}
\end{figure}

\begin{figure}[tb] %%%%%%%%%%%%
\includegraphics[width=\figwidth]{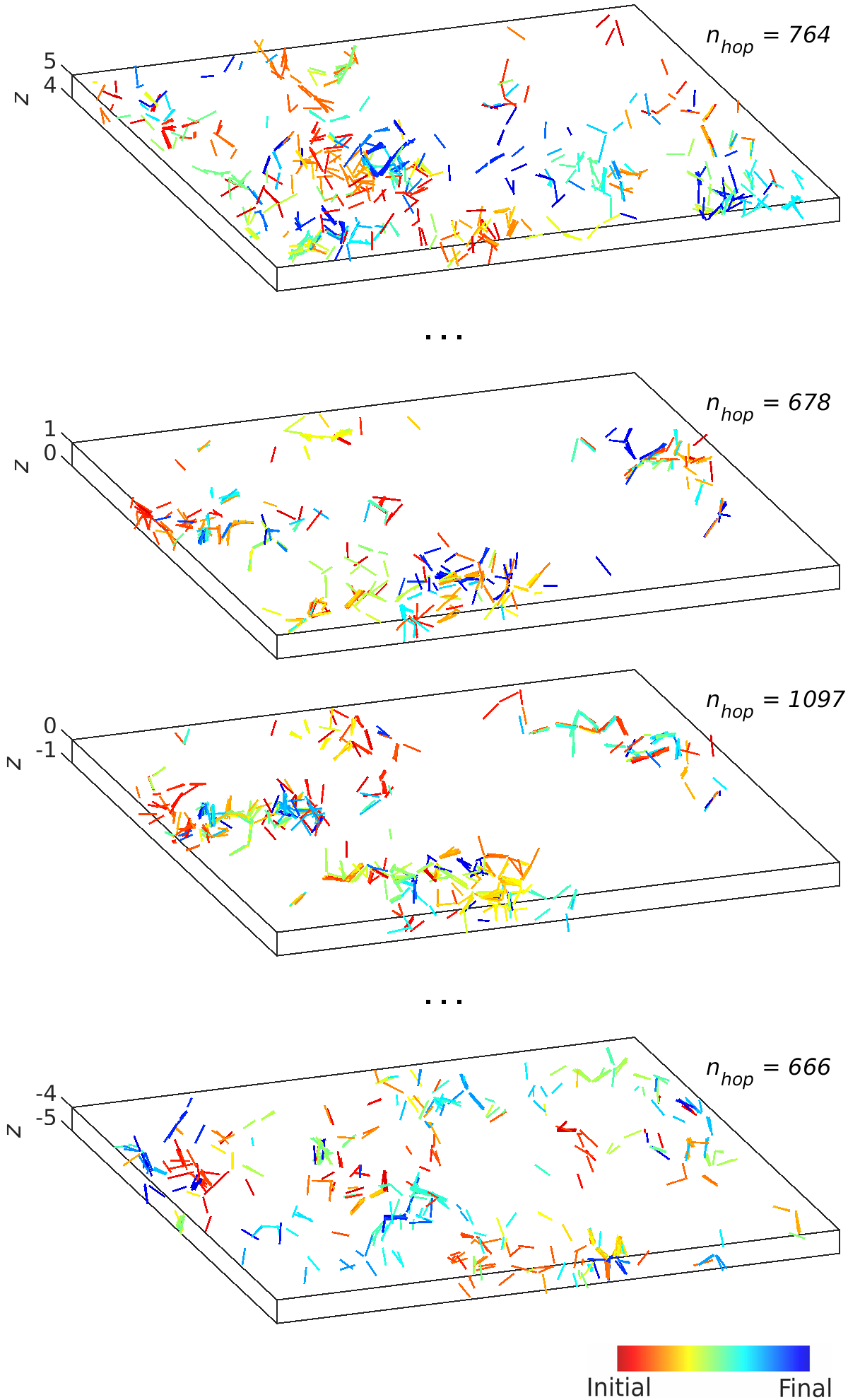}
\caption{ 3D views of hopping events in individual layers from \fig{Fblock}(b) each of thickness $\sigma$.   Stronger dynamic heterogeneity is observed at  $\abs{z} \in [0,\sigma]$ (middle layers) than at $\abs{z} \in [4\sigma,5\sigma]$  (top and bottom layers), although the numbers of hops $n_{hop}$ are similar as dictated by the uniform bulk-like hopping rate $R_1$.  
}
\label{Flayer}
\end{figure}

Particle motions in glassy systems are known to exhibit strong spatial-temporal correlations in the form of dynamic heterogeneity \cite{berthier2011book}. The surface mobile layer is expected to show reduced dynamic heterogeneity because of its Arrhenius dynamics characteristic of non-glassy liquids \cite{tsui2010}. 
In this section, we show qualitatively that surface effects reducing the dynamic heterogeneity penetrate into the inner-surface layer. Quantitative analysis will be presented in the next section.

Discretizing time by defining $t_k=k\taumin$. 
The displacement $\drik$ of particle $i$ at time $t_k$ during $\taumin$ is defined by
\begin{equation}
\label{drik}
  \drik = { \abs{\r_i(t_{k+1})- \r_i(t_k) } },
\end{equation}
which is the short-time particular case of \eq{dri}. 
Particle $i$ is considered as having hopped at time $t_k$ if $\drik\ge 0.6\sigma$.
\Fig{Fblockfull} shows all hopping events in the film during $0\le t_k < \tau$ where $\tau=10,000$. Specifically, 
if particle $i$ hops at time $t_k$, its trajectory during the hop is illustrated by a line segment joining $\r_i(t_k)$ and $\r_i(t_{k+1})$. Non-hopping parts of the trajectories are omitted.
Trajectories are colored based on the value $t_k/\tau$, so that hops at similar times are shaded in similar  colors.

We observe from \fig{Fblockfull} that hopping events are much more numerous close to the surface. This illustrates  enhanced surface mobility ultimately resulting from the reduced particle coordination at the surface. Beneath the surfaces where individual hopping events can be resolved, we observe many string-like motions \cite{glotzer1998} each corresponding to multiple particle trajectories lining up to form a nearly continuous curve typically punctuated by tiny gaps. A closer look can also reveal reversals and repetitions of strings corresponding to particle back-and-forth hopping motions, as indicated by one string closely retracing another one  \cite{lam2017}. 

Enhanced mobility propagates into the film via string-like motions originating from close to the free surfaces. 
The spatial extent of string-like motions thus provides a minimum length scale characterizing the depth variation of the particle hopping rate.
The density of hopping events in \fig{Fblockfull} is proportional to $R_1$ and   decreases monotonically with $\abs{z}$. Focusing on the region in which $R_1$ and the hopping event density have converged to their bulk values, \fig{Fblock}(a) replots the same hopping events from \fig{Fblockfull} but limited to the inner-surface and the bulk-like layers at $\abs{z} \le 5\sigma$.
Similarly, \fig{Fblock}(b) shows additional hopping events in the same region by extending the imaged period to $\tau=100,000$. 
Dynamic heterogeneity is readily observed and is depicted mainly as concentrations of hopping events in between relatively empty regions. 
It can also be observed via temporal correlations of the hops as revealed by correlations in the colors of the trajectories. 
It is evident from \fig{Fblock}(b) that surface effects reducing the dynamic heterogeneity penetrate into the inner-surface layer. This is in sharp contrast to $R_1$ which exhibits no surface effect in this layer.
Similar trends are also barely discernible in \Fig{Fblock}(a) despite stronger statistical fluctuations.

The hopping events in \fig{Fblock}(b) are resolved into layers according to the position $\r_i(t_k)$ at the beginning of a hop. Four examples of these layers are shown in \Fig{Flayer}. The numbers of hopping events $n_{hop}$ in the layers are also shown, which are basically uniform apart from statistical fluctuations. 
We observe stronger concentrations of events at $z=\pm 0.5\sigma$ at the film center. In contrast, events at $z=\pm 4.5\sigma$ are clearly more homogeneously distributed, indicating reduced heterogeneity. Surface effects on dynamic heterogeneity thus evidently extends into the inner-surface layer.

\subsection{Hopping event correlations}

\begin{figure}[tb] %%%%%%%%%%%%
\includegraphics[width=0.94\figwidth]{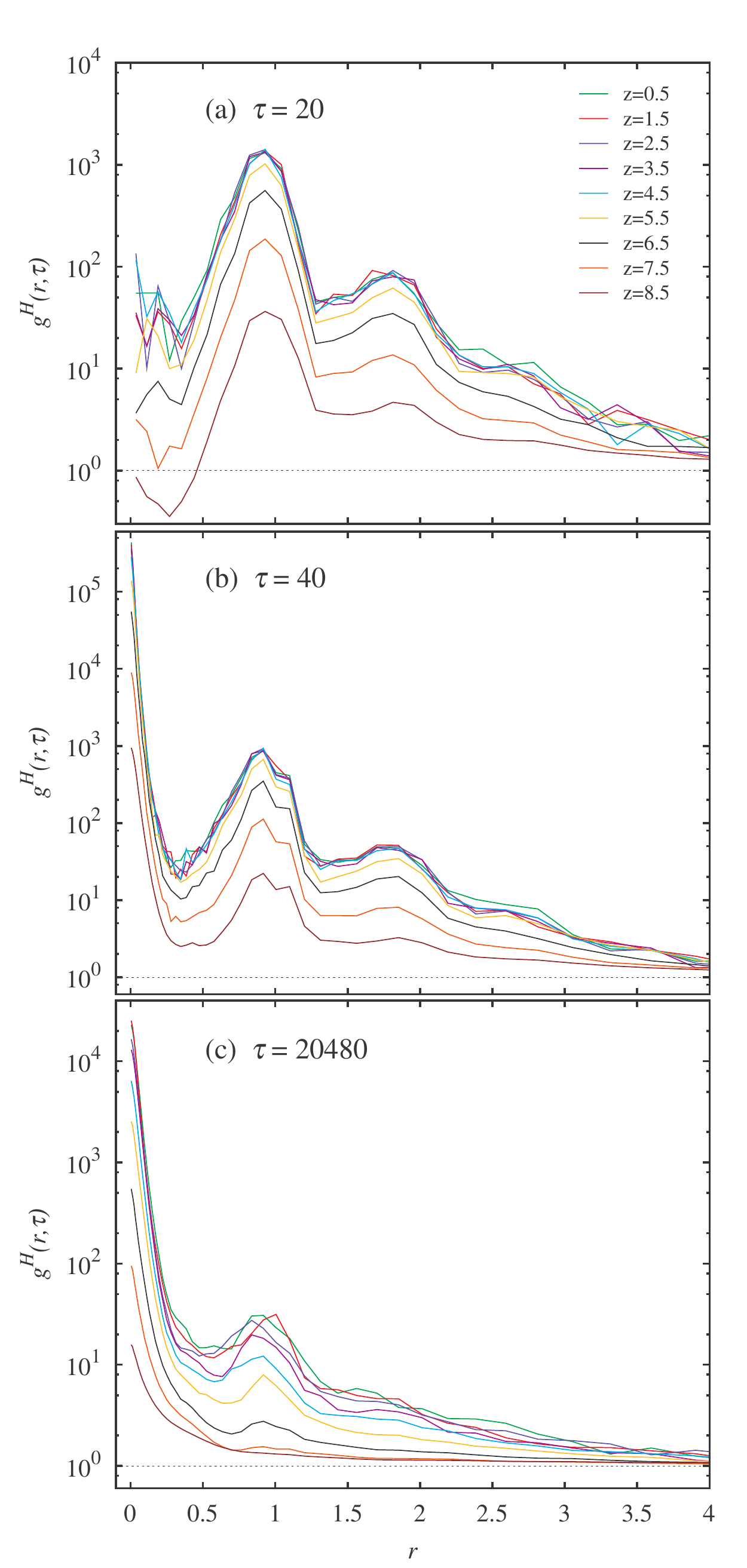}
\caption{ Hopping event pair distribution function $g^H(r,\tau)$ for hops occurring over a duration $\tau=20$ (a), 40 (b) and 20480 (c). Hops are defined based on displacements during a time $\Dtau=20$. As $z$ increases, 
$g^H(r,\tau)$ decreases and this shows surface induced reduction of dynamic heterogeneity. Surface effects have practically terminated at $z=4.5 \sigma$ in (a) and (b) but only at $z=2.5 \sigma$ in (c).
}
\label{Fhopcorr}
\end{figure}

\begin{figure}[tb] %%%%%%%%%%%%
\includegraphics[width=\figwidth]{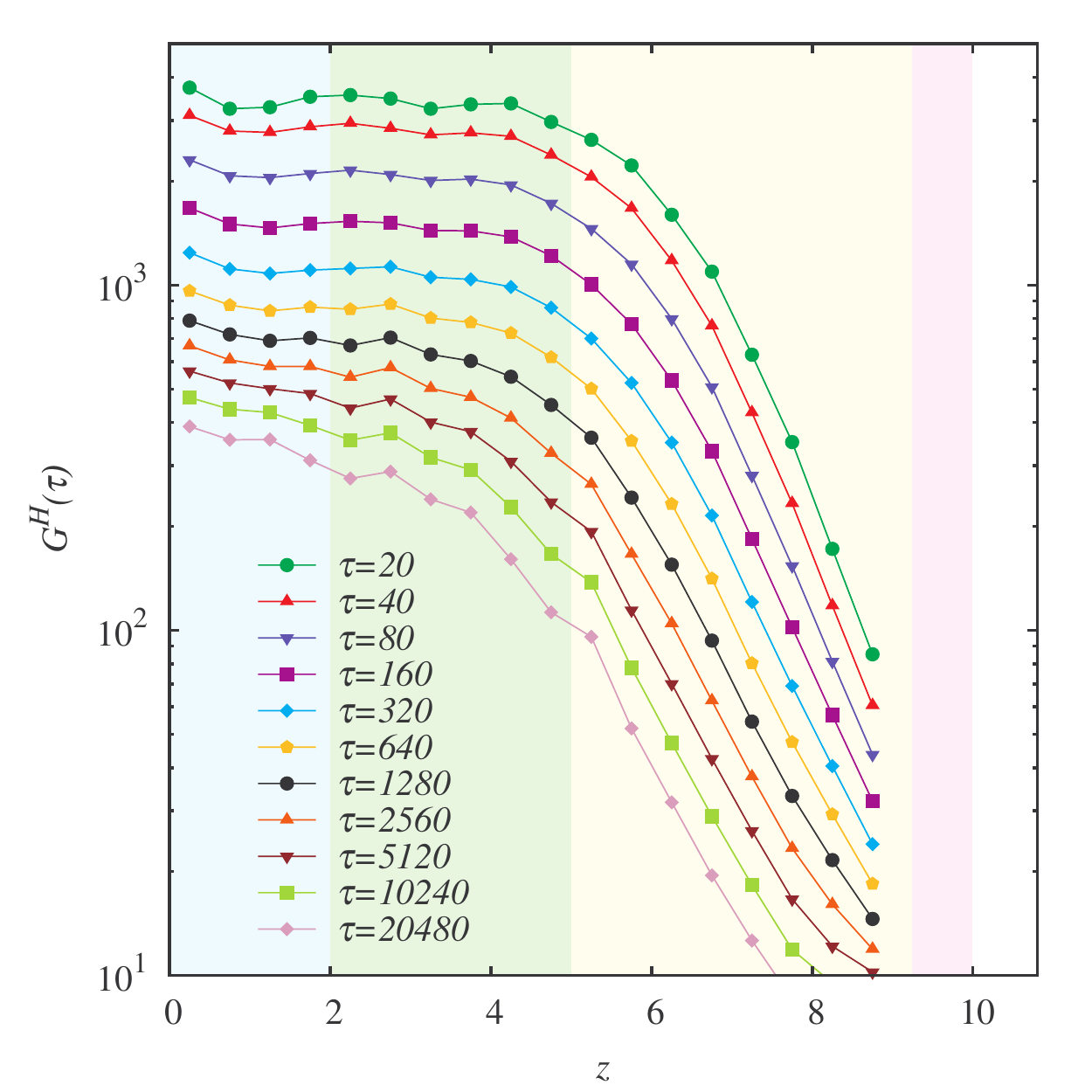}
\caption{ Correlation $G^H(\tau)$ for hopping events occurring over a duration $\tau$. 
}
\label{FhopcorrG}
\end{figure}

We now further analyze the dynamic heterogeneity quantitatively. 
We represent the location of a hopping event by the mean position
\begin{equation}
  \label{rik}
  \rik = \frac12 (\r_i(t_k) + \r_i(t_{k+1}) ).
\end{equation}
The 2D local hopping event density in layer $\Omega_z$ during time $\dtau$ is given by
\begin{equation}
  \label{denh}
 \denh(\u,\tau) =  \sum_{k=1}^{\tau/\Dtau} \sum_{i \in \Omega_z} \theta ( \drik  - 0.6 \sigma) ~ \deltatwo (\u - \u_i^{k} )    ~  
\end{equation}
where $\u_i^{k}$ denotes the projection of $\rik$ onto the $xy$-plane.
Furthermore, the 2D pair distribution function of hopping events can be defined, analogous to \eq{gr}, by  % page 30
\begin{equation}
  \label{dencorr}
  g^H(r,\tau) = \frac{ \av{ \denh(\u_0,\tau)\denh(\u_0+\u,\tau ) }} {\av{\denh(\u_0,\tau )}^2}
\end{equation}
where $r= \abs \u$ and the averages are over all 2D positions $\u_0$.  
It is numerically evaluated using a form analogous to \eq{gr2}.

\Fig{Fhopcorr}(a) plots $g^H(r,\tau)$ against $r$ for $\tau=20$ (i.e. $\taumin$). It involves only a single snapshot of hopping events 
and $g^H(r,\tau)$ is simply a layer-resolved pair distribution function of the most mobile particles. Peaks are observed at $r\simeq 0.9\sigma$, $1.7\sigma$, etc., similar to previous studies for bulk systems \cite{donati1999}. They correspond to nearest, next nearest neighbors, etc., within string-like motions. 
\Fig{Fhopcorr}(b)-(c) shows $g^H(r,\tau)$ for $\tau=40$ and 20480 (i.e. $\taumax$) respectively. A prominent main peak at $r= 0$ is also observed. It indicates abundance of multiple hops at the same position at different times $t_k$ and are mainly due to back-and-forth hopping motions. 

We observe from \fig{Fhopcorr}(a)-(c) that $g^H(r,\tau)$ is reduced close to the surface. Surface effects penetrate up to $z \simeq 5.5\sigma$  for $\tau=20$ and 40, but reach deeper to $z\simeq 4.5\sigma$ for $\tau=\taumax$. 
To establish this more clearly, we numerically evaluate an integrated hopping event correlation defined by
\begin{equation}
  \label{GH}
  G^H(\tau) = 2 \pi \int_0^{4\sigma}  r ~ \left(g^H(r,\tau)-1\right) dr
\end{equation}
motivated by $g^H(r,\tau)=1$ as $r\to \infty$.
Results are plotted in 
\fig{FhopcorrG} which further show that surface effects penetrate deeper as $\tau$ increases.
In \eq{GH}, the integration upper bound is taken as $4\sigma$ to include a large regime where $g^H(r,\tau)\gg 1$ for better statistics, but other values give qualitatively similar results.

\begin{figure}[bt] %%%%%%%%%%%%
\includegraphics[width=\figwidth]{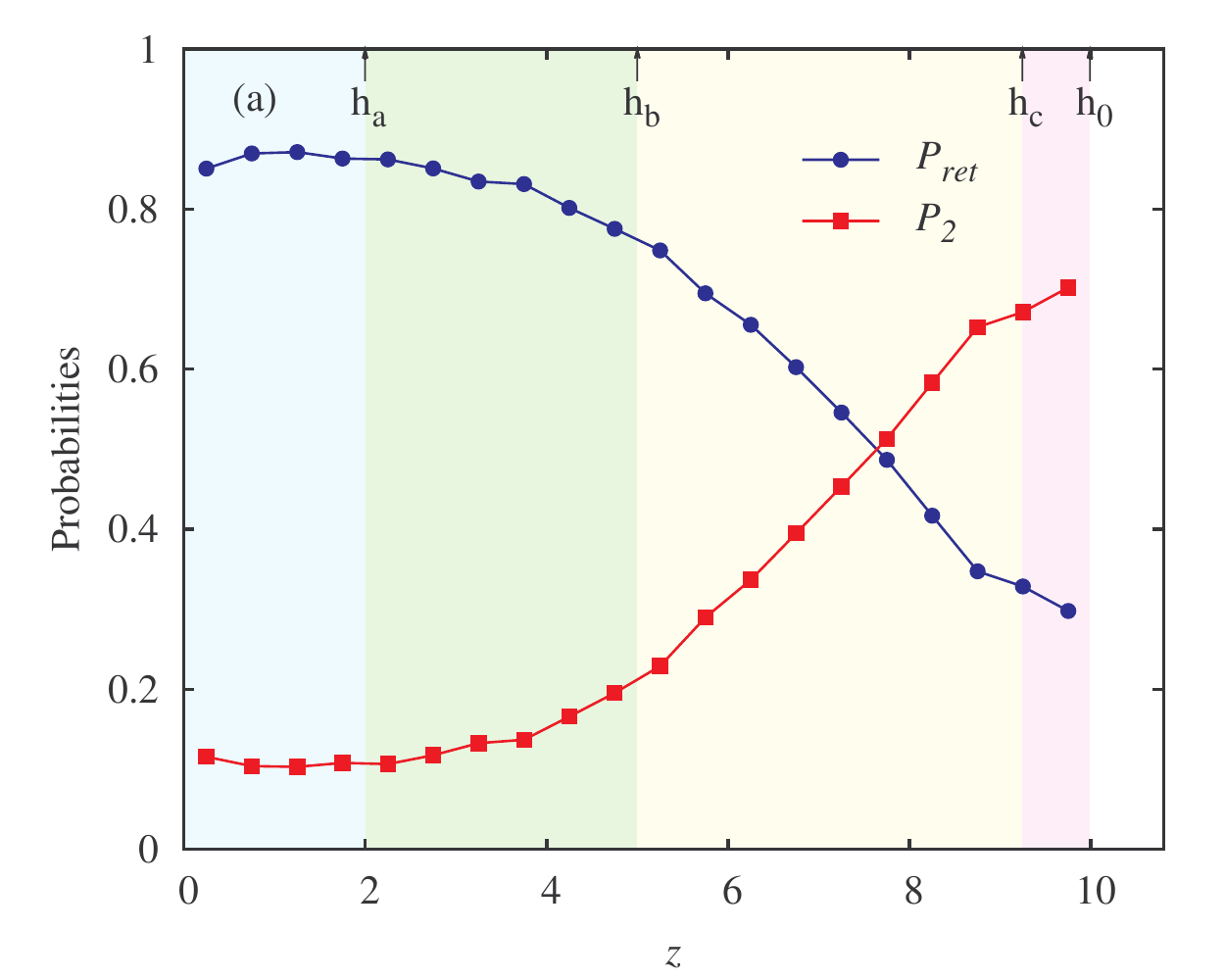}
\includegraphics[width=\figwidth]{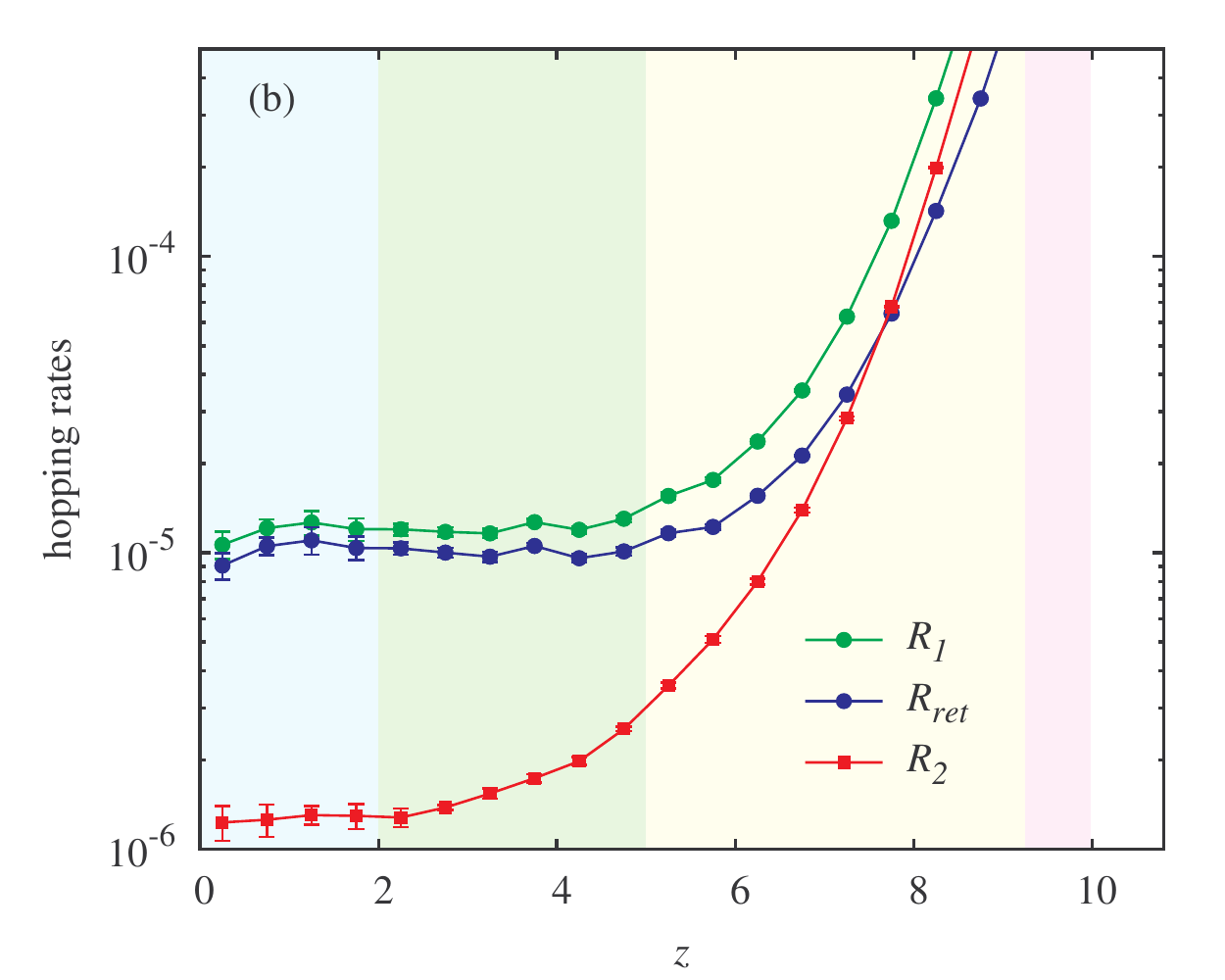}
\caption{ 
Probabilities $P_{ret}$ and $P_2$ (a) of returning and non-returning second hops and the corresponding rates  $R_{ret}$ and $R_2$ (b) against coordinate $z$. The hopping rate $R_1$ is also shown in (b).
}
\label{FPret}
\end{figure}

Particle back-and-forth motions have long been studied in glassy systems \cite{miyagawa1988,vollmayr2004,vogel2008,kawasaki2013,ahn2013,helfferich2014,yu2017,lu2016,lam2017}
 and cause the main peak at $r=0$ in \fig{Fhopcorr}(b)-(c). We now analyze them  by generalizing the approach in \Ref{lam2017} to layer-resolved measurements. Specifically, after particle $i$ in layer $\Omega_z$ has hopped at time $t_k$, we define its further motion as a returning hop if it first returns to within a distance $0.3\sigma$ from the original position $\r(t_k)$. Alternatively, the motion is defined as a non-returning second hop, i.e. an escaping hop, if it first displaces again elsewhere beyond a distance $0.6\sigma$ from the hopped position $\r(t_{k+1})$. We monitor the particle up to time 
$t_{k}+2\e5$ which is long enough 
so that the subsequent motions can be categorized in most cases. We hence calculate the probabilities $P_{ret}$ and $P_2$ that the particle first performs a returning or a non-returning second hop respectively. 
Results are shown in \fig{FPret}(a).  We observe that in the bulk-like layer, $P_{ret} \simeq 0.86$ which is a very high value implying a surprisingly strong temporal anti-correlation in the hopping of individual particles. 
In the mobile layer, $P_{ret}$ decreases monotonically towards the free surface. 
At the outer-surface layer, the much smaller value of $P_{ret} \simeq 0.30$ is consistent with simple uncorrelated motions as is expected from the non-glassy nature of the surface layer \cite{tsui2010}. 
Note that the value 0.86 in the bulk is larger than $0.73$ obtained in \Ref{lam2017} from bulk simulations because significantly more detailed trajectories with positions recorded every period $\taumin \equiv 20$ are used here so that most instances of back-and-forth motions are captured in the analysis. 

We next calculate the rates $R_{ret}$ and $R_2$ of returning and non-returning second hops using
\begin{equation}
  \label{Rret}
  R_{ret} = R_1 P_{ret} \text{ ~~and ~~~} R_2 = R_1 P_2.
\end{equation}
where $R_1$ as defined in \eq{R1} can be interpreted as the rate of the first hop, noting that every hopping event can be considered as the first of a sequence of two hops.
Since a non-returning second hop is an essential step for a large displacement of a particle, $R_2$ is closer to the structural relaxation rate and is a better characterization of the dynamics than $R_1$ as already demonstrated in \Ref{lam2017}. 
\Fig{FPret}(b) plots the measured values together with $R_1$ from \fig{FR}. 
It shows that both $R_{ret}$ and $R_2$ admit surface enhancements. However, surface effects extend to the inner-surface layer only for $R_2$ but not noticeably for $R_{ret}$. 
This again illustrates the diverse penetration depths of surface effects on different dynamical measurements and will be further discussed in the next section.

\section{Origins of the outer-, mid- and inner-surface layers}
\label{sublayer}

\begin{figure}[tb] %%%%%%%%%%%%
\includegraphics[width=\figwidth]{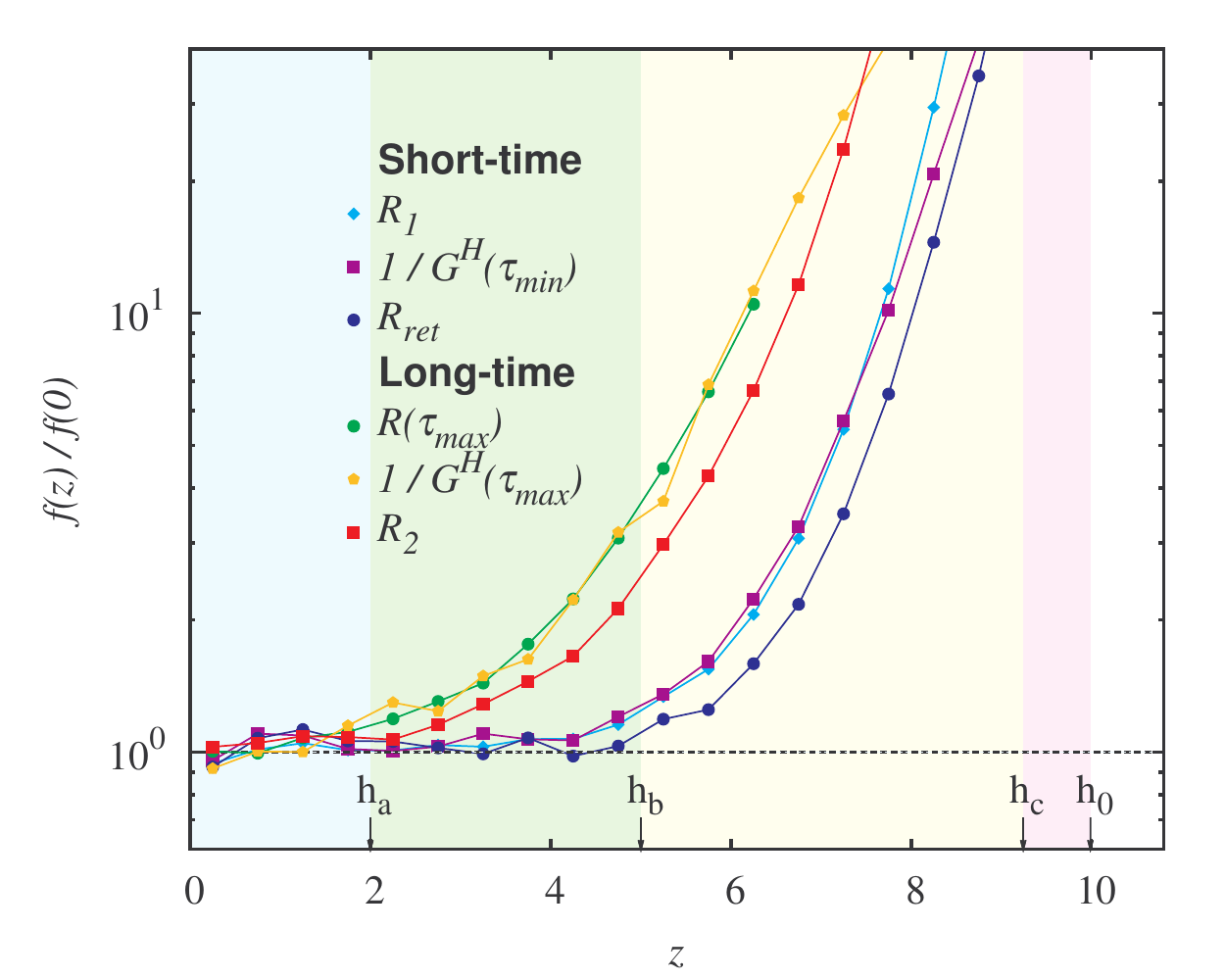}
\caption{ Layer-resolved quantity $f(z)$ normalized with respect to bulk value $f(0)$ against coordinate $z$. Quantities shown include the net hopping rate $\R$ and  the inverse correlation $1/G_H(\dtau)$  at $\taumin \equiv 20$ and $\taumax \equiv 20480$,
with $R_1 \equiv \Rmin$.
Also shown are the rates $R_{ret}$ and $R_2$ of particle returning and non-returning second hops. 
}
\label{Fall}
\end{figure}

We have defined three sublayers of the surface mobile layer, which are color-shaded in plots of quantities against $z$ in Figs. \ref{Fdensity}, \ref{Fmsd}(b), \ref{FR}, \ref{FmsdR1}, \ref{FhopcorrG}, \ref{FPret}(a)-(b) and \ref{Fall} for easy comparison.
The outer-surface layer defined in \sec{static} is characterized by a reduced density $\Den$. We expect that the density reduction is simply due to the surface roughness and particle arrangements are  already bulk-like right beneath the local position of the surface. This is supported by the good fit of $\Den$ by \eq{rhoz} motivated by surfaces limited by surface tension \cite{nelson2004}. A further support is from $g(r)$ in \fig{Fdencorr} in which the positions of the two subpeaks of the main peak coincide well with the energy minimized separations $0.96 \sigma$ and $1.12 \sigma$ of the bonded and non-bonded pair potentials. The subpeak positions remain unchanged even very close to the free surface.
Structures are thus dominated by nearest neighboring interactions and are not significantly perturbed by missing further neighbors, as next nearest neighbor interactions are much weaker.

In \sec{dynamics}, the mid-surface layer is characterized by a bulk-like $\Den$ but an enhanced hopping rate $R_1$. It can be understood qualitatively as follows. 
Particle motions at low temperature are dominated by micro-string hopping motions \cite{glotzer2004}. In each of these motions, $l$ participating particles arranged linearly hop simultaneously to displace their adjacent neighbors within the micro-string. Generalizing for convenience to include the $l=1$ case, all particle hops are considered as micro-string motions \cite{chandler2011}. They constitute more general string-like motions \cite{glotzer1998} each of which in general comprises of multiple non-simultaneous micro-strings. 
Particle hopping motions in the form of micro-string motions have been considered as elementary motions in the structural relaxations of glassy systems \cite{chandler2011,ciamarra2016,lam2017},  a 
 view consistent with potential energy landscape (PEL) and activation energy barrier calculations \cite{swayamjyoti2014}. Simultaneous hops of multiple particles in a micro-string can be favorable because the bonds between neighboring moving particles need not be broken. 

We suggest that the enhancement of the hopping rate $R_1$ at the outer- and mid-surface layers is a simple consequence of surface affects on the PEL.
A missing neighbor of a micro-string at the free surface in general alters the PEL and leads to a lower hopping energy barrier. 
Micro-string motions concerning at least one site at $z\agt h_c-\sigma$ thus admit reduced barriers. 
Interpreting $h_c-\sigma-h_b=3.25\sigma$ as the maximum lateral extent of micro-strings,
only micro-strings located completely at $z \agt h_b$ may be able to enjoy  reduced barriers and thus an enhanced $R_1$. 
The lengths of strings follow an exponential distribution with an average of about two particles long \cite{glotzer2003}. Micro-strings are their constituents and are even shorter. A maximum lateral extent of $3.25\sigma$ assumed above should be reasonable.

From \fig{FmsdR1}, the rather abrupt convergence of $R_1$ to its bulk value at $z\simeq h_b$ 
indicates that micro-string motions are very localized events depending only on the immediate neighborhood of the sites concerned. This implies that barriers based on, for instance, elastic models \cite{dyre2006review} with interactions typically decaying as power-laws may not be applicable.
In contrast, the MSD at $\taumin$ shows a much more gradual convergence to the bulk value. We suggest that this is because the MSD accounts for not only hopping but also elastic distortions. Specifically, when a particle hops, the structural perturbations can be represented by a force dipole which generates elastic distortions decaying with distance in a power-law \cite{villain1998book}. This leads to creep motions (see \sec{displacement}) of neighboring particles. Closer to the free surface,  creep motions are more significant due to the much more numerous hops. This picture is supported by the observation from \fig{FPr1} that the displacement distribution $P(\dr)$ at the inner-surface layer differs from that in the bulk-like layer mainly by having more creep motions rather than hops as explained in \sec{displacement}.

The inner-surface layer demonstrates that surface effects admit different penetration depths even for different dynamical measurements.  The contrast in the penetration depths for $R_1$ and $\Rmax$ is already demonstrated in \fig{FmsdR1}. 
As further examples of hopping related dynamics measurements,  \fig{Fall} replots $R_1$ and $\Rmax$ together with $1/G_H(\Dtau)$, $1/G_H(\taumax)$, $R_{ret}$ and $R_2$ after normalization by their bulk values.
Note that $R_{ret}$ and $R_2$ can be categorized respectively as short- and long-time measurements based on the average waiting times of the corresponding processes. From \fig{Fall}, it is interesting to observe that the normalized quantities resemble each other in the respective groups of short- and long-time measurements. Moreover, one group differs from the other mainly by a shift along the $z$ axis. Therefore, surface effects on all hopping related dynamical measurements studied in this work show distinct penetration depths in the short- and long-time regimes.

The normalized quantities in \fig{FmsdR1} are fitted to the empirical form
\begin{equation}
  \label{fit}
  \frac{f(z)}{f(0)} = 1 + \exp\left( \frac{z-z_0}{\lambda_M} \right).
\end{equation}
Here, $\lambda_M$ is a characteristic width of the surface mobile layer as probed by the quantity $f(z)$ and $z_0$ is the position at which the surface effects become significant. 
The fits are good except for the MSD at $\taumin$ as it only applies up to $f(z)/f(0) \alt 3$ compared to about 10 for the other cases.  The different behavior of the MSD at $\taumin$ is expected to be due to  elastic distortions induced by hops at the surface as explained above.
Focusing on the hopping statistics, we get $\lambda_M=0.67\sigma$ and $z_0 = 6.4 \sigma$ for $f(z)=R_1$, while $\lambda_M=0.95\sigma$ and $z_0 = 4.2 \sigma$ for $f(z)=\Rmax$. The difference between the two values of $z_0$ hence provide a more accurate estimate of $2.2\sigma$ for the thickness of the inner-surface layer, which is consistent with the thickness $2.0\sigma$ adopted above.  Other quantities shown in \fig{Fall} can also be well fitted to \eq{fit}.
\Eq{fit} can be rewritten as ${f(z)} = {f(0)} + {f(0)}\exp\left( {(z-z_0)}/{\lambda_M} \right)$.
The two terms physically account for events intrinsic to the bulk and induced by the free surface respectively. The exponential decay may be a consequence of the exponential distribution of the lengths of the strings \cite{glotzer2003}.

The exponential form followed by $\Rmax$ in \eq{fit} defines a mobility profile for long-time motions. It is expected to be the cause of a related exponential profile followed by the layer-resolved flow velocity under steady-state driven conditions reported in \Ref{lam2013crossover}.  The characteristic decay width $\lambda_M=0.95\sigma$ obtained above for $\Rmax$ indeed agrees very well with the corresponding width of $\lambda_M=0.94\sigma$ for the flow velocity profile from \Ref{lam2013crossover}. 

In the inner-surface layer, we thus observe the co-existence of a bulk-like $R_1$ with enhanced mobility.
This seemingly contradictory phenomena can be better understood based on the probabilities $P_{ret}$ and $P_2$ of returning and non-return second hops. 
From \fig{FPret}(a), a high value of $P_{ret}$ is observed in the bulk-like layer, implying a significant slowdown due to strong anti-correlations in the hopping motions. At the inner-surface layer, $P_{ret}$ is comparatively lower indicating  reduced  anti-correlations in particle hops and thus enhanced mobility.
This reduction of anti-correlations is also reflected quantitatively in $g^H$ and $G^H$ as well as visually in the dynamic heterogeneity. The enhanced mobility in the inner-surface layer hence results from diminished hopping anti-correlations rather than more frequent hops.

\section{Facilitation via diminishing hopping anti-correlations}
\label{facil}

Widely studied theories of glass include the Adam-Gibbs theory \cite{adam1965}, mode-coupling theory \cite{gotzebook}, dynamic facilitation theory \cite{fredrickson1984,palmer1984,ritort2003review,garrahan2011review,lam2017dplm,   lam2018tree}, random first order transition theory \cite{kirkpatrick1989}, elastic models \cite{dyre2006review} and so on. 
We have shown above that anti-correlations in hopping events are important in understanding surface enhanced mobility. 
Theories emphasizing the importance of correlations in elementary motions such as the facilitation picture \cite{fredrickson1984,palmer1984,ritort2003review,garrahan2011review,lam2017dplm,   lam2018tree} are most promising in describing our findings.

Dynamic facilitation often describes the phenomenon that motions in a local region can initiate other subsequent motions in a neighboring local region \cite{biroli2013review}. As is visually evident from \fig{Fblockfull}, the abundant motions close to the free surface facilitate motions deeper in the film. For the inner-surface layer where the hopping rate is already bulk-like, enhanced motions result from facilitation by the extra motions in the mid-surface layer. However, the facilitation does not increase the rate of  hopping motions in the inner-surface layer, which is essentially fixed by the bulk-like PEL. Instead, it acts by suppressing the anti-correlations between hopping events. Therefore, according to our results, dynamic facilitation is in fact the phenomenon that motions in a local region reduce the anti-correlations between motions in a neighboring local region and thus enhance structural  relaxations. 

Motivated by these findings, we have recently identified a micro-string interaction process as the dynamic facilitation mechanism consistent with the above requirements \cite{lam2017}. 
An analytical study leads to a local random configuration tree theory of glass \cite{lam2018tree} which is illustrated by explicit calculations applied to a  distinguishable particle lattice  model (DPLM) \cite{lam2017dplm}. 
In this picture, micro-string motions are initiated by quasi-voids, each of which consists of neighboring free volumes transported in whole by a micro-string motion \cite{lam2017}. At low temperature, such voids are predominately trapped by the PEL to within finite regions in the configuration space and this induces the strong anti-correlations of the particle hopping motions. A micro-string motion initiated by a void perturbs the PEL experienced by other voids, which are then momentarily untrapped or, more precisely, trapped differently. 
This thus breaks the hopping anti-correlations without generating additional micro-string motions as is required by observations in this work. At the  outer- and mid-surface layers, voids are more mobile due to surface effects on the PEL. This provides the voids in the inner-surface layer with a relatively free boundary condition at the interface to the mid-surface layer. Additional void untrapping events and enhanced dynamics thus result.

\section{Discussions}
\label{discussions}

In summary, polymer films with free surfaces are simulated and analyzed in detail. We have studied structural properties including density and particle pair distribution function, as well as dynamical properties including mean square displacement, displacement distribution, particle hopping rate, long-time net hopping rate, hopping event pair distribution function, and particle returning and non-returning hopping probabilities and rates. Surface effects on particle hopping rate are qualitatively different from those on mean square displacement and terminate abruptly when going into the film. Based on the penetration depths of surface effects on respectively the film density, 
hopping rate, and long-time net hopping rate, we define the outer-, mid- and inner-sublayers of the surface mobile layer. 
The inner-surface layer shows a bulk-like particle hopping rate but an enhanced mobility. The enhanced mobility results from  reduced temporal anti-correlations of particle hops associated with a reduced dynamic heterogeneity. The observation suggests that dynamic facilitation acts by diminishing the anti-correlations rather than enhancing the rate of elementary motions.

We have reported results at $T=0.36$, which is the lowest temperature accessible for equilibrium simulations. Smaller scale simulations at higher $T$ and non-equilibrium simulations at lower $T$ indicate that as $T$ decreases, the net hopping rate $\R$ at the surface drops more mildly than in the bulk. Surface enhancement of the mobility thus increases. The exponential decay in \eq{fit} however admits a slightly reduced characteristic width $\lambda_M$. Overall, results are qualitatively similar to those reported above and there is only a weak $T$ dependence of the mobile layer thickness. 
In this work, we have studied short-chain polymer melts in this work. However, we expect that the diverse penetration depths of surface effects and the peculiar  properties of the inner-surface layer may also be qualitatively applicable to other glassy systems with dynamics dominated by particle hops. Further studies on these systems will be of great interest.

\appendix

\section{Layer resolution schemes}
\label{layercriteria}

\begin{figure}[b] %%%%%%%%%%%%
\includegraphics[width=\figwidth]{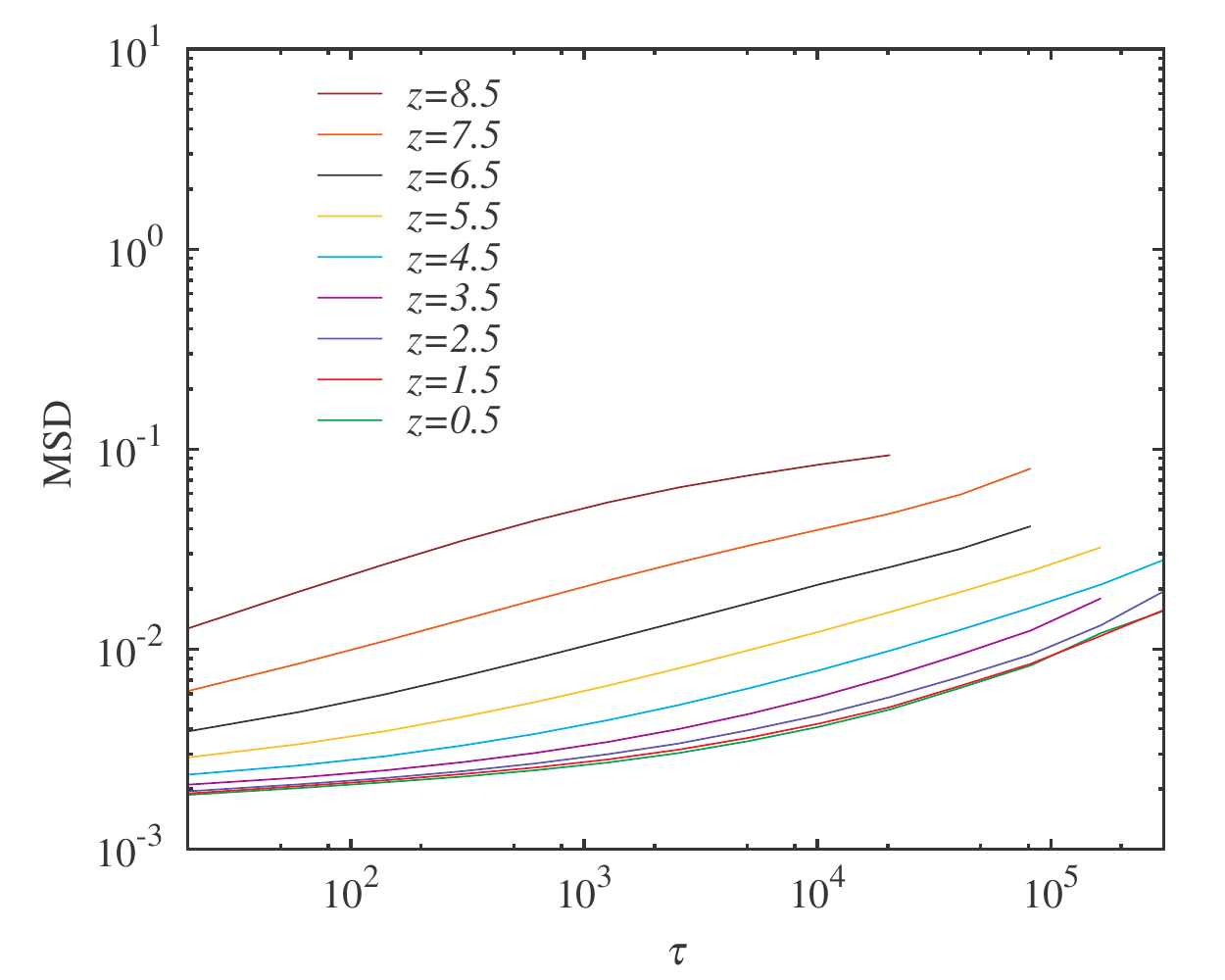}
\caption{ Mean square displacement (MSD) similar to that shown in \fig{Fmsd}(a) except for different layer-resolution criteria. Particles considered must be inside layer $\Omega_z$ at all time during the period $\dtau$. 
}
\label{FmsdS}
\end{figure}

When performing layer-resolved dynamical measurements concerning the displacement ${\r_i(t_0+\dtau)- \r_i(t_0)}$ in \eqs{dri}{drik}, we assume that particle $i$ is in layer   $\Omega_z$ solely based on its initial position $\r_i(t_0)$ at the beginning of the duration $\dtau$. This provides good statistics, consistency with bulk values, and convenience in possible  analytical calculations in the future. Since the final position $\r_i(t_0+\dtau)$ may be at a neighboring layer, this scheme in principle may provide only limited sharpness in the layer-resolution. However, we have checked that adopting two other more stringent layer-resolution criteria does not alter our results qualitatively. A main reason is that we focus mainly on hopping statistics and onset of surface perturbations concerning in most cases rather small displacements.

Specifically, we have also considered $i$ in $\Omega_z$  only if both the initial and the final positions $\r_i(t_0)$ and $\r_i(t_0+\dtau)$ are in $\Omega_z$. The resulting layer-resolved MSD is \rvs{similar to that in \fig{Fmsd}(a)}. Alternatively, we consider $i$ in $\Omega_z$ only if $\r_i(t)$ during the whole period (i.e. $t_0 \le t < t_0+\dtau$) is in $\Omega_z$, up to a time resolution limited by our recorded trajectories. This is very similar to the approach used in \Ref{baschnagel2007}. The MSD hence obtained is shown in \fig{FmsdS}. The statistics nevertheless deteriorate since the sample sizes are much reduced. Yet, compared with that in \fig{Fmsd}, values are similar when the MSD is small. More importantly, the penetration depths of the surface effects are similar. 

Adopting again the condition that both the initial and the final positions must be in $\Omega_z$, we have also calculated the net particle hopping rate $R(\dtau)$ and \rvs{the hopping event correlation $G^H(\dtau)$. Both sets of results are qualitatively similar to those in \figs{FR}{FhopcorrG} respectively} and the validity of the simple layer-resolving algorithm adopted in the main text is readily verified.

~

{\noindent\bf Acknowledgments }

~

We thank Ophelia Tsui, Fathollah Varnik, J\"{o}rg baschnagel, Simone Napolitano and Patrick Charbonneau for helpful discussions. We are grateful to the support of Hong Kong GRF (Grant 15330516).

\FloatBarrier

\bibliography{polymer_short}

%Merlin.mbs v4.21 2009-07-09.
% Polymer PRE %%%%%%%%%%%%%%%%%%%%%%%%%%%%%%%%%%%%%%%%%%%%%%% % Stringbulk
  %%%%%%%%%%%%%%%%%%%%%%%%%%%%%%%%%%%%%%%%%%%%%%%%%%%% % UGC15
  %%%%%%%%%%%%%%%%%%%%%%%%%%%%%%%%%%%%%%%%%%%%%%%%%%%% % DPLM
  %%%%%%%%%%%%%%%%%%%%%%%%%%%%%%%%%%%%%%%%%%%%%%%%%%%% % facil
  %%%%%%%%%%%%%%%%%%%%%%%%%%%%%%%%%%%%%%%%%%%%%%%%%%%% % surface
  %%%%%%%%%%%%%%%%%%%%%%%%%%%%%%%%%%%%%%%%%%%%%%%%%%%% % dplm-eq
  %%%%%%%%%%%%%%%%%%%%%%%%%%%%%%%%%%%%%%%%%%%%%%%%%%%%
\begin{thebibliography}{10}%
\makeatletter
\providecommand \@ifxundefined [1]{%
 \ifx #1\undefined \expandafter \@firstoftwo
 \else \expandafter \@secondoftwo
\fi
}%
\providecommand \@ifnum [1]{%
 \ifnum #1\expandafter \@firstoftwo
 \else \expandafter \@secondoftwo
\fi
}%
\providecommand \enquote [1]{``#1''}%
\providecommand \bibnamefont  [1]{#1}%
\providecommand \bibfnamefont [1]{#1}%
\providecommand \citenamefont [1]{#1}%
\providecommand\href[0]{\@sanitize\@href}%
\providecommand\@href[1]{\endgroup\@@startlink{#1}\endgroup\@@href}%
\providecommand\@@href[1]{#1\@@endlink}%
\providecommand \@sanitize [0]{\begingroup\catcode`\&12\catcode`\#12\relax}%
\@ifxundefined \pdfoutput {\@firstoftwo}{%
 \@ifnum{\z@=\pdfoutput}{\@firstoftwo}{\@secondoftwo}%
}{%
 \providecommand\@@startlink[1]{\leavevmode\special{html:<a href="#1">}}%
 \providecommand\@@endlink[0]{\special{html:</a>}}%
}{%
 \providecommand\@@startlink[1]{%
  \leavevmode
  \pdfstartlink
   attr{/Border[0 0 1 ]/H/I/C[0 1 1]}%
   user{/Subtype/Link/A<</Type/Action/S/URI/URI(#1)>>}%
  \relax
 }%
 \providecommand\@@endlink[0]{\pdfendlink}%
}%
\providecommand \url  [0]{\begingroup\@sanitize \@url }%
\providecommand \@url [1]{\endgroup\@href {#1}{\urlprefix}}%
\providecommand \urlprefix [0]{URL }%
\providecommand \Eprint[0]{\href }%
\@ifxundefined \urlstyle {%
  \providecommand \doi [1]{doi:\discretionary{}{}{}#1}%
}{%
  \providecommand \doi [0]{doi:\discretionary{}{}{}\begingroup
  \urlstyle{rm}\Url }%
}%
\providecommand \doibase [0]{http://dx.doi.org/}%
\providecommand \Doi[1]{\href{\doibase#1}}%
\providecommand \bibAnnote [3]{%
  \BibitemShut{#1}%
  \begin{quotation}\noindent
    \textsc{Key:}\ #2\\\textsc{Annotation:}\ #3%
  \end{quotation}%
}%
\providecommand \bibAnnoteFile [2]{%
  \IfFileExists{#2}{\bibAnnote {#1} {#2} {\input{#2}}}{}%
}%
\providecommand \typeout [0]{\immediate \write \m@ne }%
\providecommand \selectlanguage [0]{\@gobble}%
\providecommand \bibinfo [0]{\@secondoftwo}%
\providecommand \bibfield [0]{\@secondoftwo}%
\providecommand \translation [1]{[#1]}%
\providecommand \BibitemOpen[0]{}%
\providecommand \bibitemStop [0]{}%
\providecommand \bibitemNoStop [0]{.\EOS\space}%
\providecommand \EOS [0]{\spacefactor3000\relax}%
\providecommand \BibitemShut [1]{\csname bibitem#1\endcsname}%
%</preamble>
\bibitem{donthbook}%
  \BibitemOpen
  \bibfield{author}{%
  \bibinfo {author} {\bibfnamefont{E.}~\bibnamefont{Donth}},\ }%
  \emph{\bibinfo {title} {The glass transition: relaxation dynamics in liquids
  and disordered materials}},\ Vol.~\bibinfo {volume} {48}\ (\bibinfo
  {publisher} {Springer},\ \bibinfo {year} {2001})%
  \bibAnnoteFile{NoStop}{donthbook}%
\bibitem{binderbook}%
  \BibitemOpen
  \bibfield{author}{%
  \bibinfo {author} {\bibfnamefont{K.}~\bibnamefont{Binder}}\ and\ \bibinfo
  {author} {\bibfnamefont{W.}~\bibnamefont{Kob}},\ }%
  \emph{\bibinfo {title} {Glassy materials and disordered solids: An
  introduction to their statistical mechanics}}\ (\bibinfo {publisher} {World
  Scientific},\ \bibinfo {year} {2011})%
  \bibAnnoteFile{NoStop}{binderbook}%
\bibitem{biroli2013review}%
  \BibitemOpen
  \bibfield{author}{%
  \bibinfo {author} {\bibfnamefont{G.}~\bibnamefont{Biroli}}\ and\ \bibinfo
  {author} {\bibfnamefont{J.~P.}\ \bibnamefont{Garrahan}},\ }%
  \bibfield{title}{%
  \enquote{\bibinfo {title} {Perspective: The glass transition},}\ }%
  \bibfield{journal}{%
  \bibinfo {journal} {J. Chem. Phys.}\ }%
  \textbf{\bibinfo {volume} {138}},\ \bibinfo {pages} {12A301} (\bibinfo {year}
  {2013})%
  \bibAnnoteFile{NoStop}{biroli2013review}%
\bibitem{stillinger2013review}%
  \BibitemOpen
  \bibfield{author}{%
  \bibinfo {author} {\bibfnamefont{F.~H.}\ \bibnamefont{Stillinger}}\ and\
  \bibinfo {author} {\bibfnamefont{P.~G.}\ \bibnamefont{Debenedetti}},\ }%
  \bibfield{title}{%
  \enquote{\bibinfo {title} {Glass transition thermodynamics and kinetics},}\
  }%
  \bibfield{journal}{%
  \bibinfo {journal} {Annu. Rev. Condens. Matter Phys.}\ }%
  \textbf{\bibinfo {volume} {4}},\ \bibinfo {pages} {263} (\bibinfo {year}
  {2013})%
  \bibAnnoteFile{NoStop}{stillinger2013review}%
\bibitem{ediger2012review}%
  \BibitemOpen
  \bibfield{author}{%
  \bibinfo {author} {\bibfnamefont{M.~D.}\ \bibnamefont{Ediger}}\ and\ \bibinfo
  {author} {\bibfnamefont{P.}~\bibnamefont{Harrowell}},\ }%
  \bibfield{title}{%
  \enquote{\bibinfo {title} {Perspective: Supercooled liquids and glasses},}\
  }%
  \bibfield{journal}{%
  \bibinfo {journal} {J. Chem. Phys.}\ }%
  \textbf{\bibinfo {volume} {137}},\ \bibinfo {pages} {080901} (\bibinfo {year}
  {2012})%
  \bibAnnoteFile{NoStop}{ediger2012review}%
\bibitem{ediger2013review}%
  \BibitemOpen
  \bibfield{author}{%
  \bibinfo {author} {\bibfnamefont{M.~D.}\ \bibnamefont{Ediger}}\ and\ \bibinfo
  {author} {\bibfnamefont{J.~A.}\ \bibnamefont{Forrest}},\ }%
  \bibfield{title}{%
  \enquote{\bibinfo {title} {Dynamics near free surfaces and the glass
  transition in thin polymer films: a view to the future},}\ }%
  \bibfield{journal}{%
  \bibinfo {journal} {Macromolecules}\ }%
  \textbf{\bibinfo {volume} {47}},\ \bibinfo {pages} {471} (\bibinfo {year}
  {2013})%
  \bibAnnoteFile{NoStop}{ediger2013review}%
\bibitem{tsui2014review}%
  \BibitemOpen
  \bibfield{author}{%
  \bibinfo {author} {\bibfnamefont{F.}~\bibnamefont{Chen}}, \bibinfo {author}
  {\bibfnamefont{C.-H.}\ \bibnamefont{Lam}},\ and\ \bibinfo {author}
  {\bibfnamefont{O.~K.~C.}\ \bibnamefont{Tsui}},\ }%
  \bibfield{title}{%
  \enquote{\bibinfo {title} {The surface mobility of glasses},}\ }%
  \bibfield{journal}{%
  \bibinfo {journal} {Science}\ }%
  \textbf{\bibinfo {volume} {343}},\ \bibinfo {pages} {975} (\bibinfo {year}
  {2014})%
  \bibAnnoteFile{NoStop}{tsui2014review}%
\bibitem{napolitano2017}%
  \BibitemOpen
  \bibfield{author}{%
  \bibinfo {author} {\bibfnamefont{S.}~\bibnamefont{Napolitano}}, \bibinfo
  {author} {\bibfnamefont{E.}~\bibnamefont{Glynos}},\ and\ \bibinfo {author}
  {\bibfnamefont{N.~B.}\ \bibnamefont{Tito}},\ }%
  \bibfield{title}{%
  \enquote{\bibinfo {title} {Glass transition of polymers in bulk, confined
  geometries, and near interfaces},}\ }%
  \bibfield{journal}{%
  \bibinfo {journal} {Rep. Prog. Phys.}\ }%
  \textbf{\bibinfo {volume} {80}},\ \bibinfo {pages} {036602} (\bibinfo {year}
  {2017})%
  \bibAnnoteFile{NoStop}{napolitano2017}%
\bibitem{keddie1994}%
  \BibitemOpen
  \bibfield{author}{%
  \bibinfo {author} {\bibfnamefont{J.~L.}\ \bibnamefont{Keddie}}, \bibinfo
  {author} {\bibfnamefont{R.~A.~L.}\ \bibnamefont{Jones}},\ and\ \bibinfo
  {author} {\bibfnamefont{R.~A.}\ \bibnamefont{Cory}},\ }%
  \bibfield{title}{%
  \enquote{\bibinfo {title} {Size-dependent depression of the glass transition
  temperature in polymer films},}\ }%
  \bibfield{journal}{%
  \bibinfo {journal} {Europhys. Lett.}\ }%
  \textbf{\bibinfo {volume} {27}},\ \bibinfo {pages} {59} (\bibinfo {year}
  {1994})%
  \bibAnnoteFile{NoStop}{keddie1994}%
\bibitem{kawana2001}%
  \BibitemOpen
  \bibfield{author}{%
  \bibinfo {author} {\bibfnamefont{S.}~\bibnamefont{Kawana}}\ and\ \bibinfo
  {author} {\bibfnamefont{R.~A.~L.}\ \bibnamefont{Jones}},\ }%
  \bibfield{title}{%
  \enquote{\bibinfo {title} {Character of the glass transition in thin
  supported polymer films},}\ }%
  \bibfield{journal}{%
  \bibinfo {journal} {Phys. Rev. E}\ }%
  \textbf{\bibinfo {volume} {63}},\ \bibinfo {pages} {021501} (\bibinfo {year}
  {2001})%
  \bibAnnoteFile{NoStop}{kawana2001}%
\bibitem{baschnagel2006}%
  \BibitemOpen
  \bibfield{author}{%
  \bibinfo {author} {\bibfnamefont{S.}~\bibnamefont{Peter}}, \bibinfo {author}
  {\bibfnamefont{H.}~\bibnamefont{Meyer}},\ and\ \bibinfo {author}
  {\bibfnamefont{J.}~\bibnamefont{Baschnagel}},\ }%
  \bibfield{title}{%
  \enquote{\bibinfo {title} {Thickness-dependent reduction of the
  glass-transition temperature in thin polymer films with a free surface},}\ }%
  \bibfield{journal}{%
  \bibinfo {journal} {J. Polym. Phys. B}\ }%
  \textbf{\bibinfo {volume} {44}},\ \bibinfo {pages} {2951} (\bibinfo {year}
  {2006})%
  \bibAnnoteFile{NoStop}{baschnagel2006}%
\bibitem{baschnagel2007}%
  \BibitemOpen
  \bibfield{author}{%
  \bibinfo {author} {\bibfnamefont{S.}~\bibnamefont{Peter}}, \bibinfo {author}
  {\bibfnamefont{H.}~\bibnamefont{Meyer}}, \bibinfo {author}
  {\bibfnamefont{J.}~\bibnamefont{Baschnagel}},\ and\ \bibinfo {author}
  {\bibfnamefont{R.}~\bibnamefont{Seemann}},\ }%
  \bibfield{title}{%
  \enquote{\bibinfo {title} {Slow dynamics and glass transition in simulated
  free-standing polymer films: a possible relation between global and local
  glass transition temperatures},}\ }%
  \bibfield{journal}{%
  \bibinfo {journal} {J. Phys. Condens. Matter}\ }%
  \textbf{\bibinfo {volume} {19}},\ \bibinfo {pages} {205119} (\bibinfo {year}
  {2007})%
  \bibAnnoteFile{NoStop}{baschnagel2007}%
\bibitem{tsui2010}%
  \BibitemOpen
  \bibfield{author}{%
  \bibinfo {author} {\bibfnamefont{Z.}~\bibnamefont{Yang}}, \bibinfo {author}
  {\bibfnamefont{Y.}~\bibnamefont{Fujii}}, \bibinfo {author}
  {\bibfnamefont{F.~K.}\ \bibnamefont{Lee}}, \bibinfo {author}
  {\bibfnamefont{C.~H.}\ \bibnamefont{Lam}},\ and\ \bibinfo {author}
  {\bibfnamefont{O.~K.~C.}\ \bibnamefont{Tsui}},\ }%
  \bibfield{title}{%
  \enquote{\bibinfo {title} {Glass transition dynamics and surface layer
  mobility in unentangled polystyrene films},}\ }%
  \bibfield{journal}{%
  \bibinfo {journal} {Science}\ }%
  \textbf{\bibinfo {volume} {328}},\ \bibinfo {pages} {1676} (\bibinfo {year}
  {2010})%
  \bibAnnoteFile{NoStop}{tsui2010}%
\bibitem{ediger2011}%
  \BibitemOpen
  \bibfield{author}{%
  \bibinfo {author} {\bibfnamefont{L.}~\bibnamefont{Zhu}}, \bibinfo {author}
  {\bibfnamefont{C.~W.}\ \bibnamefont{Brian}}, \bibinfo {author}
  {\bibfnamefont{S.~F.}\ \bibnamefont{Swallen}}, \bibinfo {author}
  {\bibfnamefont{P.~T.}\ \bibnamefont{Straus}}, \bibinfo {author}
  {\bibfnamefont{M.~D.}\ \bibnamefont{Ediger}},\ and\ \bibinfo {author}
  {\bibfnamefont{L.}~\bibnamefont{Yu}},\ }%
  \bibfield{title}{%
  \enquote{\bibinfo {title} {Surface self-diffusion of an organic glass},}\ }%
  \bibfield{journal}{%
  \bibinfo {journal} {Phys. Rev. Lett.}\ }%
  \textbf{\bibinfo {volume} {106}},\ \bibinfo {pages} {256103} (\bibinfo {year}
  {2011})%
  \bibAnnoteFile{NoStop}{ediger2011}%
\bibitem{forrest2014}%
  \BibitemOpen
  \bibfield{author}{%
  \bibinfo {author} {\bibfnamefont{Y.}~\bibnamefont{Chai}}, \bibinfo {author}
  {\bibfnamefont{T.}~\bibnamefont{Salez}}, \bibinfo {author}
  {\bibfnamefont{J.~D.}\ \bibnamefont{McGraw}}, \bibinfo {author}
  {\bibfnamefont{M.}~\bibnamefont{Benzaquen}}, \bibinfo {author}
  {\bibfnamefont{K.}~\bibnamefont{Dalnoki-Veress}}, \bibinfo {author}
  {\bibfnamefont{E.}~\bibnamefont{Rapha{\"e}l}},\ and\ \bibinfo {author}
  {\bibfnamefont{J.~A.}\ \bibnamefont{Forrest}},\ }%
  \bibfield{title}{%
  \enquote{\bibinfo {title} {A direct quantitative measure of surface mobility
  in a glassy polymer},}\ }%
  \bibfield{journal}{%
  \bibinfo {journal} {Science}\ }%
  \textbf{\bibinfo {volume} {343}},\ \bibinfo {pages} {994} (\bibinfo {year}
  {2014})%
  \bibAnnoteFile{NoStop}{forrest2014}%
\bibitem{tanaka1998}%
  \BibitemOpen
  \bibfield{author}{%
  \bibinfo {author} {\bibfnamefont{T.}~\bibnamefont{Kajiyama}}, \bibinfo
  {author} {\bibfnamefont{K.}~\bibnamefont{Tanaka}}, \bibinfo {author}
  {\bibfnamefont{N.}~\bibnamefont{Satomi}},\ and\ \bibinfo {author}
  {\bibfnamefont{A.}~\bibnamefont{Takahara}},\ }%
  \bibfield{title}{%
  \enquote{\bibinfo {title} {Surface relaxation process of monodisperse
  polystyrene film based on lateral force microscopic measurements},}\ }%
  \bibfield{journal}{%
  \bibinfo {journal} {Macromolecules}\ }%
  \textbf{\bibinfo {volume} {31}},\ \bibinfo {pages} {5150} (\bibinfo {year}
  {1998})%
  \bibAnnoteFile{NoStop}{tanaka1998}%
\bibitem{forrest2008}%
  \BibitemOpen
  \bibfield{author}{%
  \bibinfo {author} {\bibfnamefont{Z.}~\bibnamefont{Fakhraai}}\ and\ \bibinfo
  {author} {\bibfnamefont{J.~A.}\ \bibnamefont{Forrest}},\ }%
  \bibfield{title}{%
  \enquote{\bibinfo {title} {Measuring the surface dynamics of glassy
  polymers},}\ }%
  \bibfield{journal}{%
  \bibinfo {journal} {Science}\ }%
  \textbf{\bibinfo {volume} {319}},\ \bibinfo {pages} {600} (\bibinfo {year}
  {2008})%
  \bibAnnoteFile{NoStop}{forrest2008}%
\bibitem{herminghaus2002}%
  \BibitemOpen
  \bibfield{author}{%
  \bibinfo {author} {\bibfnamefont{S.}~\bibnamefont{Herminghaus}},\ }%
  \bibfield{title}{%
  \enquote{\bibinfo {title} {Polymer thin films and surfaces: Possible effects
  of capillary waves},}\ }%
  \bibfield{journal}{%
  \bibinfo {journal} {Eur. Phys. J. E}\ }%
  \textbf{\bibinfo {volume} {8}},\ \bibinfo {pages} {237} (\bibinfo {year}
  {2002})%
  \bibAnnoteFile{NoStop}{herminghaus2002}%
\bibitem{long2001}%
  \BibitemOpen
  \bibfield{author}{%
  \bibinfo {author} {\bibfnamefont{D.}~\bibnamefont{Long}}\ and\ \bibinfo
  {author} {\bibfnamefont{F.}~\bibnamefont{Lequeux}},\ }%
  \bibfield{title}{%
  \enquote{\bibinfo {title} {Heterogeneous dynamics at the glass transition in
  van der waals liquids, in the bulk and in thin films},}\ }%
  \bibfield{journal}{%
  \bibinfo {journal} {Eur. Phys. J. E}\ }%
  \textbf{\bibinfo {volume} {4}},\ \bibinfo {pages} {371} (\bibinfo {year}
  {2001})%
  \bibAnnoteFile{NoStop}{long2001}%
\bibitem{lipson2010}%
  \BibitemOpen
  \bibfield{author}{%
  \bibinfo {author} {\bibfnamefont{S.~T.}\ \bibnamefont{Milner}}\ and\ \bibinfo
  {author} {\bibfnamefont{J.~E.~G.}\ \bibnamefont{Lipson}},\ }%
  \bibfield{title}{%
  \enquote{\bibinfo {title} {Delayed glassification model for free-surface
  suppression of t g in polymer glasses},}\ }%
  \bibfield{journal}{%
  \bibinfo {journal} {Macromolecules}\ }%
  \textbf{\bibinfo {volume} {43}},\ \bibinfo {pages} {9865} (\bibinfo {year}
  {2010})%
  \bibAnnoteFile{NoStop}{lipson2010}%
\bibitem{starr2014}%
  \BibitemOpen
  \bibfield{author}{%
  \bibinfo {author} {\bibfnamefont{P.~Z.}\ \bibnamefont{Hanakata}}, \bibinfo
  {author} {\bibfnamefont{J.~F.}\ \bibnamefont{Douglas}},\ and\ \bibinfo
  {author} {\bibfnamefont{F.~W.}\ \bibnamefont{Starr}},\ }%
  \bibfield{title}{%
  \enquote{\bibinfo {title} {Interfacial mobility scale determines the scale of
  collective motion and relaxation rate in polymer films},}\ }%
  \bibfield{journal}{%
  \bibinfo {journal} {Nat. Comm.}\ }%
  \textbf{\bibinfo {volume} {5}},\ \bibinfo {pages} {4163} (\bibinfo {year}
  {2014})%
  \bibAnnoteFile{NoStop}{starr2014}%
\bibitem{forrest2015string}%
  \BibitemOpen
  \bibfield{author}{%
  \bibinfo {author} {\bibfnamefont{T.}~\bibnamefont{Salez}}, \bibinfo {author}
  {\bibfnamefont{J.}~\bibnamefont{Salez}}, \bibinfo {author}
  {\bibfnamefont{K.}~\bibnamefont{Dalnoki-Veress}}, \bibinfo {author}
  {\bibfnamefont{E.}~\bibnamefont{Rapha{\"e}l}},\ and\ \bibinfo {author}
  {\bibfnamefont{J.~A.}\ \bibnamefont{Forrest}},\ }%
  \bibfield{title}{%
  \enquote{\bibinfo {title} {Cooperative strings and glassy interfaces},}\ }%
  \bibfield{journal}{%
  \bibinfo {journal} {Proc. Natl. Acad. Sci.}\ }%
  \textbf{\bibinfo {volume} {112}},\ \bibinfo {pages} {8227} (\bibinfo {year}
  {2015})%
  \bibAnnoteFile{NoStop}{forrest2015string}%
\bibitem{paeng2011}%
  \BibitemOpen
  \bibfield{author}{%
  \bibinfo {author} {\bibfnamefont{K.}~\bibnamefont{Paeng}}, \bibinfo {author}
  {\bibfnamefont{S.~F.}\ \bibnamefont{Swallen}},\ and\ \bibinfo {author}
  {\bibfnamefont{M.~D.}\ \bibnamefont{Ediger}},\ }%
  \bibfield{title}{%
  \enquote{\bibinfo {title} {Direct measurement of molecular motion in
  freestanding polystyrene thin films},}\ }%
  \bibfield{journal}{%
  \bibinfo {journal} {J. Am. Chem. Soc.}\ }%
  \textbf{\bibinfo {volume} {133}},\ \bibinfo {pages} {8444} (\bibinfo {year}
  {2011})%
  \bibAnnoteFile{NoStop}{paeng2011}%
\bibitem{tsui2013pmma}%
  \BibitemOpen
  \bibfield{author}{%
  \bibinfo {author} {\bibfnamefont{R.~N.}\ \bibnamefont{Li}}, \bibinfo {author}
  {\bibfnamefont{F.}~\bibnamefont{Chen}}, \bibinfo {author}
  {\bibfnamefont{C.-H.}\ \bibnamefont{Lam}},\ and\ \bibinfo {author}
  {\bibfnamefont{O.~K.~C.}\ \bibnamefont{Tsui}},\ }%
  \bibfield{title}{%
  \enquote{\bibinfo {title} {Viscosity of pmma on silica: Epitome of systems
  with strong polymer--substrate interactions},}\ }%
  \bibfield{journal}{%
  \bibinfo {journal} {Macromolecules}\ }%
  \textbf{\bibinfo {volume} {46}},\ \bibinfo {pages} {7889} (\bibinfo {year}
  {2013})%
  \bibAnnoteFile{NoStop}{tsui2013pmma}%
\bibitem{roth2010}%
  \BibitemOpen
  \bibfield{author}{%
  \bibinfo {author} {\bibfnamefont{J.~E.}\ \bibnamefont{Pye}}, \bibinfo
  {author} {\bibfnamefont{K.~A.}\ \bibnamefont{Rohald}}, \bibinfo {author}
  {\bibfnamefont{E.~A.}\ \bibnamefont{Baker}},\ and\ \bibinfo {author}
  {\bibfnamefont{C.~B.}\ \bibnamefont{Roth}},\ }%
  \bibfield{title}{%
  \enquote{\bibinfo {title} {Physical aging in ultrathin polystyrene films:
  Evidence of a gradient in dynamics at the free surface and its connection to
  the glass transition temperature reductions},}\ }%
  \bibfield{journal}{%
  \bibinfo {journal} {Macromolecules}\ }%
  \textbf{\bibinfo {volume} {43}},\ \bibinfo {pages} {8296} (\bibinfo {year}
  {2010})%
  \bibAnnoteFile{NoStop}{roth2010}%
\bibitem{ogieglo2018}%
  \BibitemOpen
  \bibfield{author}{%
  \bibinfo {author} {\bibfnamefont{W.}~\bibnamefont{Ogieglo}}, \bibinfo
  {author} {\bibfnamefont{K.}~\bibnamefont{Tempelman}}, \bibinfo {author}
  {\bibfnamefont{S.}~\bibnamefont{Napolitano}},\ and\ \bibinfo {author}
  {\bibfnamefont{N.~E.}\ \bibnamefont{Benes}},\ }%
  \bibfield{title}{%
  \enquote{\bibinfo {title} {Evidence of a transition layer between the free
  surface and the bulk},}\ }%
  \bibfield{journal}{%
  \bibinfo {journal} {J. Phys. Chem. letters}\ }%
  \textbf{\bibinfo {volume} {9}},\ \bibinfo {pages} {1195} (\bibinfo {year}
  {2018})%
  \bibAnnoteFile{NoStop}{ogieglo2018}%
\bibitem{mckenzie2018}%
  \BibitemOpen
  \bibfield{author}{%
  \bibinfo {author} {\bibfnamefont{I.}~\bibnamefont{McKenzie}}, \bibinfo
  {author} {\bibfnamefont{Y.}~\bibnamefont{Chai}}, \bibinfo {author}
  {\bibfnamefont{D.~L.}\ \bibnamefont{Cortie}}, \bibinfo {author}
  {\bibfnamefont{J.~A.}\ \bibnamefont{Forrest}}, \bibinfo {author}
  {\bibfnamefont{D.}~\bibnamefont{Fujimoto}}, \bibinfo {author}
  {\bibfnamefont{V.~L.}\ \bibnamefont{Karner}}, \bibinfo {author}
  {\bibfnamefont{R.~F.}\ \bibnamefont{Kiefl}}, \emph{et~al.},\ }%
  \bibfield{title}{%
  \enquote{\bibinfo {title} {Direct measurements of the temperature, depth and
  processing dependence of phenyl ring dynamics in polystyrene thin films by
  $\beta$-detected {NMR}},}\ }%
  \bibfield{journal}{%
  \bibinfo {journal} {Soft Matter}}%
   (\bibinfo {year} {2018})%
  \bibAnnoteFile{NoStop}{mckenzie2018}%
\bibitem{jain2004}%
  \BibitemOpen
  \bibfield{author}{%
  \bibinfo {author} {\bibfnamefont{T.~S.}\ \bibnamefont{Jain}}\ and\ \bibinfo
  {author} {\bibfnamefont{J.~J.}\ \bibnamefont{de~Pablo}},\ }%
  \bibfield{title}{%
  \enquote{\bibinfo {title} {Investigation of transition states in bulk and
  freestanding film polymer glasses},}\ }%
  \bibfield{journal}{%
  \bibinfo {journal} {Phys. Rev. Lett.}\ }%
  \textbf{\bibinfo {volume} {92}},\ \bibinfo {pages} {155505} (\bibinfo {year}
  {2004})%
  \bibAnnoteFile{NoStop}{jain2004}%
\bibitem{starr2012}%
  \BibitemOpen
  \bibfield{author}{%
  \bibinfo {author} {\bibfnamefont{P.~Z.}\ \bibnamefont{Hanakata}}, \bibinfo
  {author} {\bibfnamefont{J.~F.}\ \bibnamefont{Douglas}},\ and\ \bibinfo
  {author} {\bibfnamefont{F.~W.}\ \bibnamefont{Starr}},\ }%
  \bibfield{title}{%
  \enquote{\bibinfo {title} {Local variation of fragility and glass transition
  temperature of ultra-thin supported polymer ﬁlms},}\ }%
  \bibfield{journal}{%
  \bibinfo {journal} {J. Chem. Phys.}\ }%
  \textbf{\bibinfo {volume} {137}},\ \bibinfo {pages} {244901} (\bibinfo {year}
  {2012})%
  \bibAnnoteFile{NoStop}{starr2012}%
\bibitem{lam2013crossover}%
  \BibitemOpen
  \bibfield{author}{%
  \bibinfo {author} {\bibfnamefont{C.-H.}\ \bibnamefont{Lam}}\ and\ \bibinfo
  {author} {\bibfnamefont{O.~K.~C.}\ \bibnamefont{Tsui}},\ }%
  \bibfield{title}{%
  \enquote{\bibinfo {title} {Crossover to surface flow in supercooled
  unentangled polymer films},}\ }%
  \bibfield{journal}{%
  \bibinfo {journal} {Phys. Rev. E}\ }%
  \textbf{\bibinfo {volume} {88}},\ \bibinfo {pages} {042604} (\bibinfo {year}
  {2013})%
  \bibAnnoteFile{NoStop}{lam2013crossover}%
\bibitem{kremer1990}%
  \BibitemOpen
  \bibfield{author}{%
  \bibinfo {author} {\bibfnamefont{K.}~\bibnamefont{Kremer}}\ and\ \bibinfo
  {author} {\bibfnamefont{G.~S.}\ \bibnamefont{Grest}},\ }%
  \bibfield{title}{%
  \enquote{\bibinfo {title} {Dynamics of entangled linear polymer melts: A
  molecular-dynamics simulation},}\ }%
  \bibfield{journal}{%
  \bibinfo {journal} {J. Chem. Phys.}\ }%
  \textbf{\bibinfo {volume} {92}},\ \bibinfo {pages} {5057} (\bibinfo {year}
  {1990})%
  \bibAnnoteFile{NoStop}{kremer1990}%
\bibitem{varnik2002pre}%
  \BibitemOpen
  \bibfield{author}{%
  \bibinfo {author} {\bibfnamefont{F.}~\bibnamefont{Varnik}}, \bibinfo {author}
  {\bibfnamefont{J.}~\bibnamefont{Baschnagel}},\ and\ \bibinfo {author}
  {\bibfnamefont{K.}~\bibnamefont{Binder}},\ }%
  \bibfield{title}{%
  \enquote{\bibinfo {title} {Reduction of the glass transition temperature in
  polymer films: A molecular-dynamics study},}\ }%
  \bibfield{journal}{%
  \bibinfo {journal} {Phys. Rev. E}\ }%
  \textbf{\bibinfo {volume} {65}},\ \bibinfo {pages} {021507} (\bibinfo {year}
  {2002})%
  \bibAnnoteFile{NoStop}{varnik2002pre}%
\bibitem{varnik2002}%
  \BibitemOpen
  \bibfield{author}{%
  \bibinfo {author} {\bibfnamefont{F.}~\bibnamefont{Varnik}}\ and\ \bibinfo
  {author} {\bibfnamefont{K.}~\bibnamefont{Binder}},\ }%
  \bibfield{title}{%
  \enquote{\bibinfo {title} {Shear viscosity of a supercooled polymer melt via
  nonequilibrium molecular dynamics simulations},}\ }%
  \bibfield{journal}{%
  \bibinfo {journal} {J. Chem. Phys.}\ }%
  \textbf{\bibinfo {volume} {117}},\ \bibinfo {pages} {6336} (\bibinfo {year}
  {2002})%
  \bibAnnoteFile{NoStop}{varnik2002}%
\bibitem{kremer2003}%
  \BibitemOpen
  \bibfield{author}{%
  \bibinfo {author} {\bibfnamefont{R.}~\bibnamefont{Auhl}}, \bibinfo {author}
  {\bibfnamefont{R.}~\bibnamefont{Everaers}}, \bibinfo {author}
  {\bibfnamefont{G.~S.}\ \bibnamefont{Grest}}, \bibinfo {author}
  {\bibfnamefont{K.}~\bibnamefont{Kremer}},\ and\ \bibinfo {author}
  {\bibfnamefont{S.~J.}\ \bibnamefont{Plimpton}},\ }%
  \bibfield{title}{%
  \enquote{\bibinfo {title} {Equilibration of long chain polymer melts in
  computer simulations},}\ }%
  \bibfield{journal}{%
  \bibinfo {journal} {J. Chem. Phys.}\ }%
  \textbf{\bibinfo {volume} {119}},\ \bibinfo {pages} {12718} (\bibinfo {year}
  {2003})%
  \bibAnnoteFile{NoStop}{kremer2003}%
\bibitem{scheidler2004}%
  \BibitemOpen
  \bibfield{author}{%
  \bibinfo {author} {\bibfnamefont{P.}~\bibnamefont{Scheidler}}, \bibinfo
  {author} {\bibfnamefont{W.}~\bibnamefont{Kob}},\ and\ \bibinfo {author}
  {\bibfnamefont{K.}~\bibnamefont{Binder}},\ }%
  \bibfield{title}{%
  \enquote{\bibinfo {title} {The relaxation dynamics of a supercooled liquid
  confined by rough walls},}\ }%
  \bibfield{journal}{%
  \bibinfo {journal} {J. Phys. Chem. B}\ }%
  \textbf{\bibinfo {volume} {108}},\ \bibinfo {pages} {6673} (\bibinfo {year}
  {2004})%
  \bibAnnoteFile{NoStop}{scheidler2004}%
\bibitem{baljon2005}%
  \BibitemOpen
  \bibfield{author}{%
  \bibinfo {author} {\bibfnamefont{A.~R.~C.}\ \bibnamefont{Baljon}}, \bibinfo
  {author} {\bibfnamefont{M.~H. M.~Weert}\ \bibnamefont{Van}}, \bibinfo
  {author} {\bibfnamefont{R.~B.}\ \bibnamefont{DeGraaff}},\ and\ \bibinfo
  {author} {\bibfnamefont{R.}~\bibnamefont{Khare}},\ }%
  \bibfield{title}{%
  \enquote{\bibinfo {title} {Glass transition behavior of polymer films of
  nanoscopic dimensions},}\ }%
  \bibfield{journal}{%
  \bibinfo {journal} {Macromolecules}\ }%
  \textbf{\bibinfo {volume} {38}},\ \bibinfo {pages} {2391} (\bibinfo {year}
  {2005})%
  \bibAnnoteFile{NoStop}{baljon2005}%
\bibitem{doi2006}%
  \BibitemOpen
  \bibfield{author}{%
  \bibinfo {author} {\bibfnamefont{H.}~\bibnamefont{Morita}}, \bibinfo {author}
  {\bibfnamefont{K.}~\bibnamefont{Tanaka}}, \bibinfo {author}
  {\bibfnamefont{T.}~\bibnamefont{Kajiyama}}, \bibinfo {author}
  {\bibfnamefont{T.}~\bibnamefont{Nishi}},\ and\ \bibinfo {author}
  {\bibfnamefont{M.}~\bibnamefont{Doi}},\ }%
  \bibfield{title}{%
  \enquote{\bibinfo {title} {Study of the glass transition temperature of
  polymer surface by coarse-grained molecular dynamics simulation},}\ }%
  \bibfield{journal}{%
  \bibinfo {journal} {Macromolecules}\ }%
  \textbf{\bibinfo {volume} {39}},\ \bibinfo {pages} {6233} (\bibinfo {year}
  {2006})%
  \bibAnnoteFile{NoStop}{doi2006}%
\bibitem{lam2017}%
  \BibitemOpen
  \bibfield{author}{%
  \bibinfo {author} {\bibfnamefont{C.-H.}\ \bibnamefont{Lam}},\ }%
  \bibfield{title}{%
  \enquote{\bibinfo {title} {Repetition and pair-interaction of string-like
  hopping motions in glassy polymers},}\ }%
  \bibfield{journal}{%
  \bibinfo {journal} {J. Chem. Phys.}\ }%
  \textbf{\bibinfo {volume} {146}},\ \bibinfo {pages} {244906} (\bibinfo {year}
  {2017})%
  \bibAnnoteFile{NoStop}{lam2017}%
\bibitem{hou2010}%
  \BibitemOpen
  \bibfield{author}{%
  \bibinfo {author} {\bibfnamefont{J.-X.}\ \bibnamefont{Hou}}, \bibinfo
  {author} {\bibfnamefont{C.}~\bibnamefont{Svaneborg}}, \bibinfo {author}
  {\bibfnamefont{R.}~\bibnamefont{Everaers}},\ and\ \bibinfo {author}
  {\bibfnamefont{G.~S.}\ \bibnamefont{Grest}},\ }%
  \bibfield{title}{%
  \enquote{\bibinfo {title} {Stress relaxation in entangled polymer melts},}\
  }%
  \bibfield{journal}{%
  \bibinfo {journal} {Phys. Rev. Lett.}\ }%
  \textbf{\bibinfo {volume} {105}},\ \bibinfo {pages} {068301} (\bibinfo {year}
  {2010})%
  \bibAnnoteFile{NoStop}{hou2010}%
\bibitem{edwardsbook}%
  \BibitemOpen
  \bibfield{author}{%
  \bibinfo {author} {\bibfnamefont{M.}~\bibnamefont{Doi}}\ and\ \bibinfo
  {author} {\bibfnamefont{S.~F.}\ \bibnamefont{Edwards}},\ }%
  \emph{\bibinfo {title} {The theory of polymer dynamics}}\ (\bibinfo
  {publisher} {Oxford University Press},\ \bibinfo {year} {1986})%
  \bibAnnoteFile{NoStop}{edwardsbook}%
\bibitem{hoomd}%
  \BibitemOpen
  \bibfield{author}{%
  \bibinfo {author} {\bibfnamefont{J.~A.}\ \bibnamefont{Anderson}}, \bibinfo
  {author} {\bibfnamefont{C.~D.}\ \bibnamefont{Lorenz}},\ and\ \bibinfo
  {author} {\bibfnamefont{A.}~\bibnamefont{Travesset}},\ }%
  \bibfield{title}{%
  \enquote{\bibinfo {title} {General purpose molecular dynamics simulations
  fully implemented on graphics processing units},}\ }%
  \bibfield{journal}{%
  \bibinfo {journal} {J. Comp. Phys.}\ }%
  \textbf{\bibinfo {volume} {227}},\ \bibinfo {pages} {5342} (\bibinfo {year}
  {2008})%
  \bibAnnoteFile{NoStop}{hoomd}%
\bibitem{nelson2004}%
  \BibitemOpen
  \bibfield{author}{%
  \bibinfo {author} {\bibfnamefont{D.}~\bibnamefont{Nelson}}, \bibinfo {author}
  {\bibfnamefont{T.}~\bibnamefont{Piran}},\ and\ \bibinfo {author}
  {\bibfnamefont{S.}~\bibnamefont{Weinberg}},\ }%
  \emph{\bibinfo {title} {Statistical mechanics of membranes and surfaces}}\
  (\bibinfo {publisher} {World Scientific},\ \bibinfo {year} {2004})%
  \bibAnnoteFile{NoStop}{nelson2004}%
\bibitem{hansen1990book}%
  \BibitemOpen
  \bibfield{author}{%
  \bibinfo {author} {\bibfnamefont{J.-P.}\ \bibnamefont{Hansen}}\ and\ \bibinfo
  {author} {\bibfnamefont{I.~R.}\ \bibnamefont{McDonald}},\ }%
  \emph{\bibinfo {title} {Theory of simple liquids}}\ (\bibinfo {publisher}
  {Elsevier},\ \bibinfo {year} {1990})%
  \bibAnnoteFile{NoStop}{hansen1990book}%
\bibitem{wahnstrom1991}%
  \BibitemOpen
  \bibfield{author}{%
  \bibinfo {author} {\bibfnamefont{G{\"o}ran}\ \bibnamefont{Wahnstr{\"o}m}},\
  }%
  \bibfield{title}{%
  \enquote{\bibinfo {title} {Molecular-dynamics study of a supercooled
  two-component lennard-jones system},}\ }%
  \bibfield{journal}{%
  \bibinfo {journal} {Phys. Rev. A}\ }%
  \textbf{\bibinfo {volume} {44}},\ \bibinfo {pages} {3752} (\bibinfo {year}
  {1991})%
  \bibAnnoteFile{NoStop}{wahnstrom1991}%
\bibitem{miyagawa1988}%
  \BibitemOpen
  \bibfield{author}{%
  \bibinfo {author} {\bibfnamefont{H.}~\bibnamefont{Miyagawa}}, \bibinfo
  {author} {\bibfnamefont{Y.}~\bibnamefont{Hiwatari}}, \bibinfo {author}
  {\bibfnamefont{B.}~\bibnamefont{Bernu}},\ and\ \bibinfo {author}
  {\bibfnamefont{J.~P.}\ \bibnamefont{Hansen}},\ }%
  \bibfield{title}{%
  \enquote{\bibinfo {title} {Molecular dynamics study of binary soft-sphere
  mixtures: Jump motions of atoms in the glassy state},}\ }%
  \bibfield{journal}{%
  \bibinfo {journal} {J. Chem. Phys.}\ }%
  \textbf{\bibinfo {volume} {88}},\ \bibinfo {pages} {3879} (\bibinfo {year}
  {1988})%
  \bibAnnoteFile{NoStop}{miyagawa1988}%
\bibitem{vollmayr2004}%
  \BibitemOpen
  \bibfield{author}{%
  \bibinfo {author} {\bibfnamefont{K.}~\bibnamefont{Vollmayr-Lee}},\ }%
  \bibfield{title}{%
  \enquote{\bibinfo {title} {Single particle jumps in a binary lennard-jones
  system below the glass transition},}\ }%
  \bibfield{journal}{%
  \bibinfo {journal} {J. Chem. Phys.}\ }%
  \textbf{\bibinfo {volume} {121}},\ \bibinfo {pages} {4781} (\bibinfo {year}
  {2004})%
  \bibAnnoteFile{NoStop}{vollmayr2004}%
\bibitem{vogel2008}%
  \BibitemOpen
  \bibfield{author}{%
  \bibinfo {author} {\bibfnamefont{M.}~\bibnamefont{Vogel}},\ }%
  \bibfield{title}{%
  \enquote{\bibinfo {title} {Conformational and structural relaxations of poly
  (ethylene oxide) and poly (propylene oxide) melts: Molecular dynamics study
  of spatial heterogeneity, cooperativity, and correlated forward--backward
  motion},}\ }%
  \bibfield{journal}{%
  \bibinfo {journal} {Macromolecules}\ }%
  \textbf{\bibinfo {volume} {41}},\ \bibinfo {pages} {2949} (\bibinfo {year}
  {2008})%
  \bibAnnoteFile{NoStop}{vogel2008}%
\bibitem{kawasaki2013}%
  \BibitemOpen
  \bibfield{author}{%
  \bibinfo {author} {\bibfnamefont{T.}~\bibnamefont{Kawasaki}}\ and\ \bibinfo
  {author} {\bibfnamefont{A.}~\bibnamefont{Onuki}},\ }%
  \bibfield{title}{%
  \enquote{\bibinfo {title} {Slow relaxations and stringlike jump motions in
  fragile glass-forming liquids: Breakdown of the stokes-einstein relation},}\
  }%
  \bibfield{journal}{%
  \bibinfo {journal} {Phys. Rev. E}\ }%
  \textbf{\bibinfo {volume} {87}},\ \bibinfo {pages} {012312} (\bibinfo {year}
  {2013})%
  \bibAnnoteFile{NoStop}{kawasaki2013}%
\bibitem{ahn2013}%
  \BibitemOpen
  \bibfield{author}{%
  \bibinfo {author} {\bibfnamefont{J.~W.}\ \bibnamefont{Ahn}}, \bibinfo
  {author} {\bibfnamefont{B.}~\bibnamefont{Falahee}}, \bibinfo {author}
  {\bibfnamefont{C.~D.}\ \bibnamefont{Piccolo}}, \bibinfo {author}
  {\bibfnamefont{M.}~\bibnamefont{Vogel}},\ and\ \bibinfo {author}
  {\bibfnamefont{D.}~\bibnamefont{Bingemann}},\ }%
  \bibfield{title}{%
  \enquote{\bibinfo {title} {Are rare, long waiting times between rearrangement
  events responsible for the slowdown of the dynamics at the glass
  transition?}.}\ }%
  \bibfield{journal}{%
  \bibinfo {journal} {J. Chem. Phys.}\ }%
  \textbf{\bibinfo {volume} {138}},\ \bibinfo {pages} {12A527} (\bibinfo {year}
  {2013})%
  \bibAnnoteFile{NoStop}{ahn2013}%
\bibitem{helfferich2014}%
  \BibitemOpen
  \bibfield{author}{%
  \bibinfo {author} {\bibfnamefont{J.}~\bibnamefont{Helfferich}}, \bibinfo
  {author} {\bibfnamefont{F.}~\bibnamefont{Ziebert}}, \bibinfo {author}
  {\bibfnamefont{S.}~\bibnamefont{Frey}}, \bibinfo {author}
  {\bibfnamefont{H.}~\bibnamefont{Meyer}}, \bibinfo {author}
  {\bibfnamefont{J.}~\bibnamefont{Farago}}, \bibinfo {author}
  {\bibfnamefont{A.}~\bibnamefont{Blumen}},\ and\ \bibinfo {author}
  {\bibfnamefont{J.}~\bibnamefont{Baschnagel}},\ }%
  \bibfield{title}{%
  \enquote{\bibinfo {title} {Continuous-time random-walk approach to
  supercooled liquids. i. different definitions of particle jumps and their
  consequences},}\ }%
  \bibfield{journal}{%
  \bibinfo {journal} {Phys. Rev. E}\ }%
  \textbf{\bibinfo {volume} {89}},\ \bibinfo {pages} {042603} (\bibinfo {year}
  {2014})%
  \bibAnnoteFile{NoStop}{helfferich2014}%
\bibitem{yu2017}%
  \BibitemOpen
  \bibfield{author}{%
  \bibinfo {author} {\bibfnamefont{H.-B.}\ \bibnamefont{Yu}}, \bibinfo {author}
  {\bibfnamefont{R.}~\bibnamefont{Richert}},\ and\ \bibinfo {author}
  {\bibfnamefont{K.}~\bibnamefont{Samwer}},\ }%
  \bibfield{title}{%
  \enquote{\bibinfo {title} {Structural rearrangements governing
  johari-goldstein relaxations in metallic glasses},}\ }%
  \bibfield{journal}{%
  \bibinfo {journal} {Sci. Adv.}\ }%
  \textbf{\bibinfo {volume} {3}},\ \bibinfo {pages} {e1701577} (\bibinfo {year}
  {2017})%
  \bibAnnoteFile{NoStop}{yu2017}%
\bibitem{lu2016}%
  \BibitemOpen
  \bibfield{author}{%
  \bibinfo {author} {\bibfnamefont{Y.~J.}\ \bibnamefont{L{\"u}}}\ and\ \bibinfo
  {author} {\bibfnamefont{W.~H.}\ \bibnamefont{Wang}},\ }%
  \bibfield{title}{%
  \enquote{\bibinfo {title} {Single-particle dynamics near the glass transition
  of a metallic glass},}\ }%
  \bibfield{journal}{%
  \bibinfo {journal} {Phys. Rev. E}\ }%
  \textbf{\bibinfo {volume} {94}},\ \bibinfo {pages} {062611} (\bibinfo {year}
  {2016})%
  \bibAnnoteFile{NoStop}{lu2016}%
\bibitem{chandler2011}%
  \BibitemOpen
  \bibfield{author}{%
  \bibinfo {author} {\bibfnamefont{A.~S.}\ \bibnamefont{Keys}}, \bibinfo
  {author} {\bibfnamefont{L.~O.}\ \bibnamefont{Hedges}}, \bibinfo {author}
  {\bibfnamefont{J.~P.}\ \bibnamefont{Garrahan}}, \bibinfo {author}
  {\bibfnamefont{S.~C.}\ \bibnamefont{Glotzer}},\ and\ \bibinfo {author}
  {\bibfnamefont{D.}~\bibnamefont{Chandler}},\ }%
  \bibfield{title}{%
  \enquote{\bibinfo {title} {Excitations are localized and relaxation is
  hierarchical in glass-forming liquids},}\ }%
  \bibfield{journal}{%
  \bibinfo {journal} {Phys. Rev. X}\ }%
  \textbf{\bibinfo {volume} {1}},\ \bibinfo {pages} {021013} (\bibinfo {year}
  {2011})%
  \bibAnnoteFile{NoStop}{chandler2011}%
\bibitem{berthier2011book}%
  \BibitemOpen
  \bibfield{author}{%
  \bibinfo {author} {\bibfnamefont{L.}~\bibnamefont{Berthier}}, \bibinfo
  {author} {\bibfnamefont{G.}~\bibnamefont{Biroli}}, \bibinfo {author}
  {\bibfnamefont{J.-P.}\ \bibnamefont{Bouchaud}}, \bibinfo {author}
  {\bibfnamefont{L.}~\bibnamefont{Cipelletti}},\ and\ \bibinfo {author}
  {\bibfnamefont{W.}~\bibnamefont{van Saarloos}},\ }%
  \emph{\bibinfo {title} {Dynamical heterogeneities in glasses, colloids, and
  granular media}},\ Vol.\ \bibinfo {volume} {150}\ (\bibinfo {publisher}
  {Oxford University Press},\ \bibinfo {year} {2011})%
  \bibAnnoteFile{NoStop}{berthier2011book}%
\bibitem{glotzer1998}%
  \BibitemOpen
  \bibfield{author}{%
  \bibinfo {author} {\bibfnamefont{C.}~\bibnamefont{Donati}}, \bibinfo {author}
  {\bibfnamefont{J.~F.}\ \bibnamefont{Douglas}}, \bibinfo {author}
  {\bibfnamefont{W.}~\bibnamefont{Kob}}, \bibinfo {author}
  {\bibfnamefont{S.~J.}\ \bibnamefont{Plimpton}}, \bibinfo {author}
  {\bibfnamefont{P.~H.}\ \bibnamefont{Poole}},\ and\ \bibinfo {author}
  {\bibfnamefont{S.~C.}\ \bibnamefont{Glotzer}},\ }%
  \bibfield{title}{%
  \enquote{\bibinfo {title} {Stringlike cooperative motion in a supercooled
  liquid},}\ }%
  \bibfield{journal}{%
  \bibinfo {journal} {Phys. Rev. Lett.}\ }%
  \textbf{\bibinfo {volume} {80}},\ \bibinfo {pages} {2338} (\bibinfo {year}
  {1998})%
  \bibAnnoteFile{NoStop}{glotzer1998}%
\bibitem{donati1999}%
  \BibitemOpen
  \bibfield{author}{%
  \bibinfo {author} {\bibfnamefont{C.}~\bibnamefont{Donati}}, \bibinfo {author}
  {\bibfnamefont{S.~C.}\ \bibnamefont{Glotzer}}, \bibinfo {author}
  {\bibfnamefont{P.~H.}\ \bibnamefont{Poole}}, \bibinfo {author}
  {\bibfnamefont{W.}~\bibnamefont{Kob}},\ and\ \bibinfo {author}
  {\bibfnamefont{S.~J.}\ \bibnamefont{Plimpton}},\ }%
  \bibfield{title}{%
  \enquote{\bibinfo {title} {Spatial correlations of mobility and immobility in
  a glass-forming lennard-jones liquid},}\ }%
  \bibfield{journal}{%
  \bibinfo {journal} {Phys. Rev. E}\ }%
  \textbf{\bibinfo {volume} {60}},\ \bibinfo {pages} {3107} (\bibinfo {year}
  {1999})%
  \bibAnnoteFile{NoStop}{donati1999}%
\bibitem{glotzer2004}%
  \BibitemOpen
  \bibfield{author}{%
  \bibinfo {author} {\bibfnamefont{Y.}~\bibnamefont{Gebremichael}}, \bibinfo
  {author} {\bibfnamefont{M.}~\bibnamefont{Vogel}},\ and\ \bibinfo {author}
  {\bibfnamefont{S.~C.}\ \bibnamefont{Glotzer}},\ }%
  \bibfield{title}{%
  \enquote{\bibinfo {title} {Particle dynamics and the development of
  string-like motion in a simulated monoatomic supercooled liquid},}\ }%
  \bibfield{journal}{%
  \bibinfo {journal} {J. Chem. Phys.}\ }%
  \textbf{\bibinfo {volume} {120}},\ \bibinfo {pages} {4415} (\bibinfo {year}
  {2004})%
  \bibAnnoteFile{NoStop}{glotzer2004}%
\bibitem{ciamarra2016}%
  \BibitemOpen
  \bibfield{author}{%
  \bibinfo {author} {\bibfnamefont{M.~P.}\ \bibnamefont{Ciamarra}}, \bibinfo
  {author} {\bibfnamefont{R.}~\bibnamefont{Pastore}},\ and\ \bibinfo {author}
  {\bibfnamefont{A.}~\bibnamefont{Coniglio}},\ }%
  \bibfield{title}{%
  \enquote{\bibinfo {title} {Particle jumps in structural glasses},}\ }%
  \bibfield{journal}{%
  \bibinfo {journal} {Soft matter}\ }%
  \textbf{\bibinfo {volume} {12}},\ \bibinfo {pages} {358} (\bibinfo {year}
  {2016})%
  \bibAnnoteFile{NoStop}{ciamarra2016}%
\bibitem{swayamjyoti2014}%
  \BibitemOpen
  \bibfield{author}{%
  \bibinfo {author} {\bibfnamefont{S.}~\bibnamefont{Swayamjyoti}}, \bibinfo
  {author} {\bibfnamefont{J.~F.}\ \bibnamefont{L{\"o}ffler}},\ and\ \bibinfo
  {author} {\bibfnamefont{P.~M.}\ \bibnamefont{Derlet}},\ }%
  \bibfield{title}{%
  \enquote{\bibinfo {title} {Local structural excitations in model glasses},}\
  }%
  \bibfield{journal}{%
  \bibinfo {journal} {Phys. Rev. B}\ }%
  \textbf{\bibinfo {volume} {89}},\ \bibinfo {pages} {224201} (\bibinfo {year}
  {2014})%
  \bibAnnoteFile{NoStop}{swayamjyoti2014}%
\bibitem{glotzer2003}%
  \BibitemOpen
  \bibfield{author}{%
  \bibinfo {author} {\bibfnamefont{M.}~\bibnamefont{Aichele}}, \bibinfo
  {author} {\bibfnamefont{Y.}~\bibnamefont{Gebremichael}}, \bibinfo {author}
  {\bibfnamefont{F.~W.}\ \bibnamefont{Starr}}, \bibinfo {author}
  {\bibfnamefont{J.}~\bibnamefont{Baschnagel}},\ and\ \bibinfo {author}
  {\bibfnamefont{S.~C.}\ \bibnamefont{Glotzer}},\ }%
  \bibfield{title}{%
  \enquote{\bibinfo {title} {Polymer-specific effects of bulk relaxation and
  stringlike correlated motion in the dynamics of a supercooled polymer
  melt},}\ }%
  \bibfield{journal}{%
  \bibinfo {journal} {J. Chem Phys.}\ }%
  \textbf{\bibinfo {volume} {119}},\ \bibinfo {pages} {5290} (\bibinfo {year}
  {2003})%
  \bibAnnoteFile{NoStop}{glotzer2003}%
\bibitem{dyre2006review}%
  \BibitemOpen
  \bibfield{author}{%
  \bibinfo {author} {\bibfnamefont{J.~C.}\ \bibnamefont{Dyre}},\ }%
  \bibfield{title}{%
  \enquote{\bibinfo {title} {Colloquium: The glass transition and elastic
  models of glass-forming liquids},}\ }%
  \bibfield{journal}{%
  \bibinfo {journal} {Rev. Mod. Phys.}\ }%
  \textbf{\bibinfo {volume} {78}},\ \bibinfo {pages} {953} (\bibinfo {year}
  {2006})%
  \bibAnnoteFile{NoStop}{dyre2006review}%
\bibitem{villain1998book}%
  \BibitemOpen
  \bibfield{author}{%
  \bibinfo {author} {\bibfnamefont{A.}~\bibnamefont{Pimpinelli}}\ and\ \bibinfo
  {author} {\bibfnamefont{J.}~\bibnamefont{Villain}},\ }%
  \emph{\bibinfo {title} {Physics of crystal growth}},\ Vol.~\bibinfo {volume}
  {19}\ (\bibinfo {publisher} {Cambridge university press Cambridge},\ \bibinfo
  {year} {1998})%
  \bibAnnoteFile{NoStop}{villain1998book}%
\bibitem{adam1965}%
  \BibitemOpen
  \bibfield{author}{%
  \bibinfo {author} {\bibfnamefont{G.}~\bibnamefont{Adam}}\ and\ \bibinfo
  {author} {\bibfnamefont{J.~H.}\ \bibnamefont{Gibbs}},\ }%
  \bibfield{title}{%
  \enquote{\bibinfo {title} {On the temperature dependence of cooperative
  relaxation properties in glass-forming liquids},}\ }%
  \bibfield{journal}{%
  \bibinfo {journal} {J. Chem Phys.}\ }%
  \textbf{\bibinfo {volume} {43}},\ \bibinfo {pages} {139} (\bibinfo {year}
  {1965})%
  \bibAnnoteFile{NoStop}{adam1965}%
\bibitem{gotzebook}%
  \BibitemOpen
  \bibfield{author}{%
  \bibinfo {author} {\bibfnamefont{W.}~\bibnamefont{G{\H o}tze}},\ }%
  \emph{\bibinfo {title} {{Complex dynamics of glass-forming liquids: a
  mode-coupling theory}}}\ (\bibinfo {publisher} {Oxford University Press},\
  \bibinfo {year} {2008})%
  \bibAnnoteFile{NoStop}{gotzebook}%
\bibitem{fredrickson1984}%
  \BibitemOpen
  \bibfield{author}{%
  \bibinfo {author} {\bibfnamefont{G.~H.}\ \bibnamefont{Fredrickson}}\ and\
  \bibinfo {author} {\bibfnamefont{H.~C.}\ \bibnamefont{Andersen}},\ }%
  \bibfield{title}{%
  \enquote{\bibinfo {title} {Kinetic ising model of the glass transition},}\ }%
  \bibfield{journal}{%
  \bibinfo {journal} {Phys. Rev. Lett.}\ }%
  \textbf{\bibinfo {volume} {53}},\ \bibinfo {pages} {1244} (\bibinfo {year}
  {1984})%
  \bibAnnoteFile{NoStop}{fredrickson1984}%
\bibitem{palmer1984}%
  \BibitemOpen
  \bibfield{author}{%
  \bibinfo {author} {\bibfnamefont{R.~G.}\ \bibnamefont{Palmer}}, \bibinfo
  {author} {\bibfnamefont{D.~L.}\ \bibnamefont{Stein}}, \bibinfo {author}
  {\bibfnamefont{E.}~\bibnamefont{Abrahams}},\ and\ \bibinfo {author}
  {\bibfnamefont{P.~W.}\ \bibnamefont{Anderson}},\ }%
  \bibfield{title}{%
  \enquote{\bibinfo {title} {Models of hierarchically constrained dynamics for
  glassy relaxation},}\ }%
  \bibfield{journal}{%
  \bibinfo {journal} {Phys. Rev. Lett.}\ }%
  \textbf{\bibinfo {volume} {53}},\ \bibinfo {pages} {958} (\bibinfo {year}
  {1984})%
  \bibAnnoteFile{NoStop}{palmer1984}%
\bibitem{ritort2003review}%
  \BibitemOpen
  \bibfield{author}{%
  \bibinfo {author} {\bibfnamefont{F.}~\bibnamefont{Ritort}}\ and\ \bibinfo
  {author} {\bibfnamefont{P.}~\bibnamefont{Sollich}},\ }%
  \bibfield{title}{%
  \enquote{\bibinfo {title} {Glassy dynamics of kinetically constrained
  models},}\ }%
  \bibfield{journal}{%
  \bibinfo {journal} {Adv. Phys.}\ }%
  \textbf{\bibinfo {volume} {52}},\ \bibinfo {pages} {219} (\bibinfo {year}
  {2003})%
  \bibAnnoteFile{NoStop}{ritort2003review}%
\bibitem{garrahan2011review}%
  \BibitemOpen
  \bibfield{author}{%
  \bibinfo {author} {\bibfnamefont{J.~P.}\ \bibnamefont{Garrahan}}, \bibinfo
  {author} {\bibfnamefont{P.}~\bibnamefont{Sollich}},\ and\ \bibinfo {author}
  {\bibfnamefont{C.}~\bibnamefont{Toninelli}},\ }%
  \bibfield{title}{%
  \enquote{\bibinfo {title} {Kinetically constrained models},}\ }%
  \bibinfo {journal} {in Dynamical Heterogeneities in Glasses, Colloids and
  Granular Media, edited by L. Berthier, G. Biroli, J.-P. Bouchaud, L.
  Cipelletti, and W. van Saarloosand (Oxford University Press, 2011)}%
  \bibAnnoteFile{NoStop}{garrahan2011review}%
\bibitem{lam2017dplm}%
  \BibitemOpen
\bibfield{journal}{%
    }%
  \bibfield{author}{%
  \bibinfo {author} {\bibfnamefont{L.-H.}\ \bibnamefont{Zhang}}\ and\ \bibinfo
  {author} {\bibfnamefont{C.-H.}\ \bibnamefont{Lam}},\ }%
  \bibfield{title}{%
  \enquote{\bibinfo {title} {Emergent facilitation behavior in a
  distinguishable-particle lattice model of glass},}\ }%
  \bibfield{journal}{%
  \bibinfo {journal} {Phys. Rev. B}\ }%
  \textbf{\bibinfo {volume} {95}},\ \bibinfo {pages} {184202} (\bibinfo {year}
  {2017})%
  \bibAnnoteFile{NoStop}{lam2017dplm}%
\bibitem{lam2018tree}%
  \BibitemOpen
  \bibfield{author}{%
  \bibinfo {author} {\bibfnamefont{C.-H.}\ \bibnamefont{Lam}},\ }%
  \bibfield{title}{%
  \enquote{\bibinfo {title} {Local random configuration-tree theory for string
  repetition and facilitated dynamics of glass},}\ }%
  \bibfield{journal}{%
  \bibinfo {journal} {J. Stat. Mech.}\ }%
  \textbf{\bibinfo {volume} {2018}},\ \bibinfo {pages} {023301} (\bibinfo
  {year} {2018})%
  \bibAnnoteFile{NoStop}{lam2018tree}%
\bibitem{kirkpatrick1989}%
  \BibitemOpen
  \bibfield{author}{%
  \bibinfo {author} {\bibfnamefont{T.~R.}\ \bibnamefont{Kirkpatrick}}, \bibinfo
  {author} {\bibfnamefont{D.}~\bibnamefont{Thirumalai}},\ and\ \bibinfo
  {author} {\bibfnamefont{P.~G.}\ \bibnamefont{Wolynes}},\ }%
  \bibfield{title}{%
  \enquote{\bibinfo {title} {Scaling concepts for the dynamics of viscous
  liquids near an ideal glassy state},}\ }%
  \bibfield{journal}{%
  \bibinfo {journal} {Phys. Rev. A}\ }%
  \textbf{\bibinfo {volume} {40}},\ \bibinfo {pages} {1045} (\bibinfo {year}
  {1989})%
  \bibAnnoteFile{NoStop}{kirkpatrick1989}%
\end{thebibliography}%

\end{document}